\documentclass[apj,numberedappendix]{emulateapj}
\usepackage{apjfonts}
\usepackage{amsmath}
\usepackage{booktabs}
\begin{document}


%\title{Evolution of Alfv\'en Waves in Magnetar Magnetospheres and Implication in the Giant Flare}
%\title{Evolution of Magnetohydrodynamic Waves in Magnetar Magnetospheres and Implication in the Giant Flare}
\title{The Evolution of High Temperature Plasma in Magnetar Magnetospheres and its Implications for Giant Flares}

\author{Makoto Takamoto}
\affil{Max-Planck-Institut f\"ur Kernphysik, Saupfercheckweg 1, Heidelberg, D69117, Germany}
\email{makoto.takamoto@mpi-hd.mpg.de}

\author{Shota Kisaka}
%\affil{Institute for Cosmic Ray Research, University of Tokyo, 5-1-5 Kashiwa-no-ha, Kashiwa city, Chiba 277-8582, Japan}
%\email{kisaka@icrr.u-tokyo.ac.jp}
\affil{Institute of Particle and Nuclear Studies, KEK, 1-1, Oho, Tsukuba 305-0801, Japan}
\email{kisaka@post.kek.jp}

\author{Takeru Suzuki}
\affil{Department of Physics, Graduate School of Science, Nagoya University, Furo-cho, Chikusa-ku, Nagoya 464-8602, Japan}
\email{stakeru@nagoya-u.jp}

\author{Toshio Terasawa}
\affil{Institute for Cosmic Ray Research, University of Tokyo, 5-1-5 Kashiwa-no-ha, Kashiwa city, Chiba 277-8582, Japan}
\email{terasawa@icrr.u-tokyo.ac.jp}

%\author{John Kirk}
%\affil{Max-Planck-Institute fur Kernphysik, Heidelberg, Germany}
%\email{John.Kirk@mpi-hd.mpg.de}


%\altaffiltext{1}{Visiting Astronomer, Cerro Tololo Inter-American Observatory.
%CTIO is operated by AURA, Inc.\ under contract to the National Science
%\altaffiltext{2}{Society of Fellows, Harvard University.}
%\altaffiltext{3}{present address: Center for Astrophysics,
%    60 Garden Street, Cambridge, MA 02138}
%\altaffiltext{4}{Visiting Programmer, Space Telescope Science Institute}
%\altaffiltext{5}{Patron, Alonso's Bar and Grill}


\begin{abstract}
In this paper 
we propose a new mechanism describing the initial spike of giant flares in the framework of the starquake model. 
We investigate the evolution of a plasma on a closed magnetic flux tube in the magnetosphere of a magnetar in the case of a sudden energy release 
and discuss the relationship with observations of giant flares. 
We perform one-dimensional numerical simulations of the relativistic magnetohydrodynamics in Schwarzschild geometry. 
We assume energy is injected at the footpoints of the loop by a hot star surface containing random perturbations of the transverse velocity. 
Alfv\'en waves are generated and propagate upward, 
accompanying very hot plasma that is also continuously heated by nonlinearly generated compressive waves. 
We find that 
the front edges of the fireball regions collide at the top of the tube with their symmetrically launched counterparts. 
This collision results in an energy release 
which can describe the light curve of initial spikes of giant flares. 
\end{abstract}





\section{\label{sec:sec1}Introduction}

Soft gamma-ray repeaters (SGRs) and anomalous X-ray pulsars (AXPs) have been considered to be strongly magnetized astrophysical objects. 
Recent X-ray observations have shown that SGRs and AXPs have similar luminosity, 
pulse period, period derivative, and burst activity, so that they are considered to belong to the same population, 
called ``magnetars'' which are isolated neutron stars powered by the dissipation of magnetic field energy~\citep{1992ApJ...392L...9D,2006RPPh...69.2631H}. 
%Their typical period is $P \sim 10$ [s] and period derivative is $P / \dot{P} \sim 10^{3-5}$ year, 
Their typical period is $P \sim 10$ [s] and period derivative is $\dot{P} \sim 10^{-10}$ to $10^{-12}$ [s/s], 
which indicates that their magnetic field is about $B \geq 10^{13}$ [G]~\citep{2006csxs.book..547W,2008A&ARv..15..225M}. 
Concerning their X-ray spectra, 
they indicate that AXPs and SGRs have a blackbody component with temperature $k_B T \sim 0.5$ [keV]~\citep{2002ApJ...581.1280M,2005ApJ...628..938M,2006ApJ...653.1423M}, 
and $k_B T \sim 5 -10$ [keV] in the case of SGRs in their active phase~\citep{2004ApJ...612..408F,2004ApJ...616.1148O,2007PASJ...59..653N,2007A&A...476..321E}. 

%Giant flares are the sudden release of an enormous amount of energy $\sim 10^{44}$ [erg] observed from \textcolor{blue}{magnetars}. 
%Their spectra and temporal evolution consist of a short hard spike followed by a pulsating tail. 
%The initial spike is a short hard spike whose rising time is about milli-second and duration is about $0.1$ [s]. 
%\citet{2007ApJ...665L..55T} reported that 
%their total luminosity and energy of the initial spike is about $2.3 \times 10^{46}$ [erg/s] and $4.3 \times 10^{44}$ [erg], respectively, 
%and its temperature is about 240 keV, 
%which is believed to induce a electron-positron plasma outflow. 
Giant flares cause the sudden release of an enormous amount of energy $\sim 10^{44}$ [erg] 
(e.g.~\citet{2007ApJ...665L..55T}), 
and they have been observed from SGRs. % observed from magnetars. 
Their spectra and temporal evolution consist of a short hard spike followed by a pulsating tail. 
%The initial spike is a short hard spike whose rising time is about milli-second, duration is about $0.1$ [s], 
The rise-time of the spike is about one milli-second, its duration is about $0.1$ [s], 
and its temperature is about $240$ [keV]~(e.g.~\citet{2005Natur.434.1098H}), respectively. 
The tail has coherent pulsations, presumably at the spin period of the underlying neutron star. 
It has a softer spectra and gradually decays in hundreds of seconds. 
The total amount of energy of the tail is about $10^{44}$ [erg], which is believed to be the energy trapped in the magnetosphere. 

The physical mechanism of giant flares is still unclear and there are a lot of investigations
~\citep{1995MNRAS.275..255T,2002ApJ...574..332T,2003MNRAS.346..540L,2006MNRAS.367.1594L,2010PASJ...62.1093M,2013ApJ...774...92P}. 
One of the most popular mechanisms is magnetic reconnection, 
which is a process that converts magnetic field energy into the plasma thermal and kinetic energy rapidly
~\citep{2003ApJ...589..893L,2003MNRAS.346..540L,2005MNRAS.358..113L,2006MNRAS.367.1594L,2010ApJ...716L.214Z,2013ApJ...775...50T}. 
Hence, it allows to convert the strong magnetic field energy of magnetars into radiation. 
In particular, 
the twisted magnetic flux rope model \citep{2006MNRAS.367.1594L,2010MNRAS.407.1926G,2012ApJ...757...67Y,2013ApJ...771L..46Y} 
is recently considered to be a promising candidate for magnetar flare phenomena 
because the twisted component allows to trigger various kinds of plasma instabilities (e.g. \citet{2004A&A...413L..27T,2006PhRvL..96y5002K}). 
%In this paper, 
%we focused on the study of the effects of MHD waves on the dynamical evolution of magnetar magnetospheres, 
%so that we assumed a pure poloidal flux tube. 
%Extension to the twisted background flux tube is our future work. 
Although magnetic reconnection has many attractive properties, 
%magnetic reconnection itself still has uncertainties which includes its typical growing timescale and energy conversion rate into radiation. 
it is subject to uncertainties which include its typical growing timescale and its energy conversion rate into radiation 
which relate to the precise mechanism of the eruption. 
In addition, 
little is known about the global magnetic field structure in the magnetar magnetosphere
~\citep{2007Ap&SS.308..631B,2009ApJ...703.1044B,2012arXiv1209.3855T,2013ApJ...762...13B,2013ApJ...774...92P}, 
which always plays a crucial role in the formation of the current sheets needed for magnetic reconnection. 

%Concerning the flare phenomena, 
It is widely known that 
the flares are ubiquitous on the Sun. 
\citet{2010PASJ...62.1093M} pointed out that 
there are a lot of similarities between magnetar giant flares and solar flares, 
and constructed a theoretical model based on the solar flare. 
%Even though there is no conclusive observational proof that 
%\textcolor{blue}{
%the physical state of plasma in the magnetar atmosphere is similar to that in the solar atmosphere, 
%}
%we consider that 
%physical understandings of solar flares gives us a lot of fruitful implications on researches of magnetar giant flares. 
Even though there may be differences between the magnetar magnetosphere and the solar atmosphere, 
we think that 
the physical understanding of solar flares can be fruitfully applied to research on magnetar giant flares. 
Referring to numerical simulations of solar coronal loops by~\citet{2004ApJ...601L.107M}, 
transient flare-like events are frequently seen 
even though quasi-steady Alfv\'enic perturbations are injected from the footpoints of a 1D closed flux tube. 
This is interpreted as a consequence of nonlinear interaction of counter-propagating waves. 
We expect that such a mechanism possibly operates in closed loops on a magnetar. 

\begin{figure}[h]
 \centering
  \includegraphics[width=6.5cm,clip]{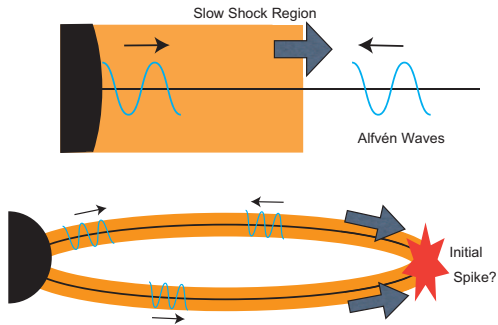}
  \caption{A schematic picture of a mechanism generating the initial spike of the giant flare 
           through the collision of slow shocks. 
           The black lines are the magnetic field line; 
           the orange regions are the post shock region of slow shocks; 
           the blue wavy lines are the Alfv\'en waves. 
          }
  \label{fig:1.1}
\end{figure}

In this paper, 
we propose a new mechanism describing the initial spike based on the starquake model
~\citep{1974Natur.251..399P,1989ApJ...343..839B,1995MNRAS.275..255T,1996ApJ...473..322T,2004ApJ...614..922K,2005ApJ...634L.153P,2008PThPh.119...39O,2011MNRAS.412.1381M,2012MNRAS.421.2054G}, 
which considers an excitation of magnetic field oscillations by the energy release of crust elastic energy through a starquake. 
We consider the evolution of fireball region along a magnetic flux tube using the general relativistic magnetohydrodynamic (GRMHD) approximation. 
In addition, we consider Alfv\'en waves driven on the star surface and study their effects on the plasma evolution. 
In our model, 
the fireball region expands along a magnetic flux tube 
and collides to the counterpart 
that propagates from the other side of the tube as indicated in Figure \ref{fig:1.1}, 
which can result in the initial spike of giant flares. 
This short timescale of the collision enables us to explain the timescale of the initial spike even in the framework of the starquake model. 
Because of the heating by Alfv\'en waves, 
the front edge of the region keeps its internal energy, 
so that the induced energy is large enough to explain the observed energy of initial spikes. 

%In this paper, we propose a new mechanism describing the initial spike. 
%We consider the evolution of fireball region along a magnetic flux tube using the general relativistic magnetohydrodynamic (GRMHD) approximation. 
%In addition, we consider Alfv\'en waves driven on the star surface and study their effects on the plasma evolution. 
%In our model, 
%the fireball region expands along a magnetic flux tube 
%and collides to the counterpart 
%that propagates from the other side of the tube as indicated in Figure \ref{fig:1.1}, 
%which can result in the initial spike of giant flares. 
%Because of the heating by Alfv\'en waves, 
%the front edge of the region keeps its internal energy, 
%so that the induced energy is large enough to explain the observed energy of initial spikes. 

This paper is organized as follows: 
In section 2, 
we explain the numerical set up and also give the basic equations. 
In section 3, 
the numerical results are presented; in addition, we present the new mechanism of the initial spike. 
In section 4, 
we give several comments on the relation to the observations.

\section{\label{sec:sec2}Method}
\subsection{\label{sec:sec2.1}Numerical Setup}

%\begin{table}[h]
%  \caption{List of Simulation Parameters}
%  \begin{tabular}{lcc} \toprule
%    Name & Amplitude $\langle \delta v / c_s \rangle$ & Radius: $[r_g / R_0]$ \\
%    \hline
%    CS05G1 & $0.5$ & $0.1$ \\
%    CS05G2 & $0.5$ & $0.2$ \\
%    CS05G3 & $0.5$ & $0.3$ \\
%    CS075G1 & $0.75$ & $0.1$ \\
%    CS07G2 & $0.75$ & $0.2$ \\
%    CS075G3 & $0.75$ & $0.3$ \\
%    CS09G1 & $0.9$ & $0.1$ \\
%    CS09G2 & $0.9$ & $0.2$ \\
%    CS09G3 & $0.9$ & $0.3$ \\
%    \hline
%  \end{tabular}
%  \label{list:1}
%\end{table}

\begin{figure*}[top]
 \centering
  \includegraphics[width=16cm,clip]{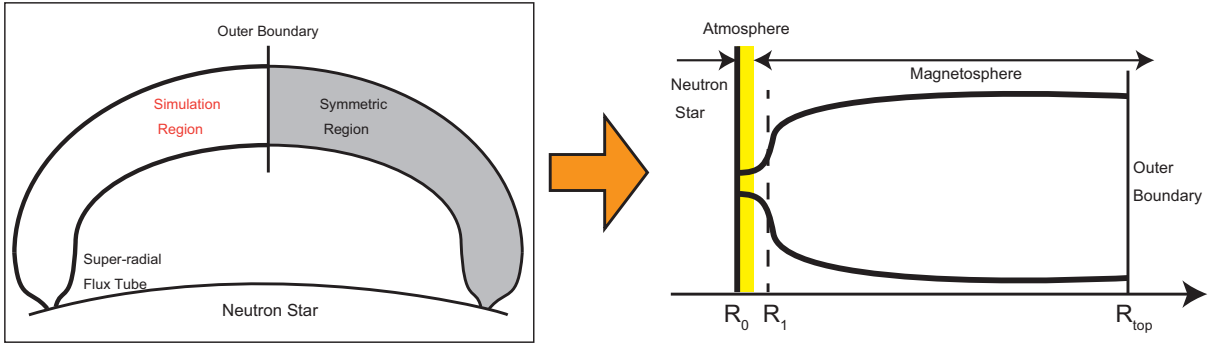}
  \caption{Schematic pictures of a flux tube with the super-radial expansion structure. 
           Left: A global picture of the flux tube. 
           Right: A local picture of the flux tube of simulation region along its central line. 
           The left-hand region is the neutron star; 
           the yellow region is the atmosphere; 
           $r = R_1$ is the coordinate where the flux tube expands mainly; 
           the right-hand region is the magnetosphere. 
          }
  \label{fig:2.1.1}
\end{figure*}

We model the evolution of a plasma in a magnetar magnetosphere %with relativistically strong magnetic field 
using the general relativistic magnetohydrodynamic (GRMHD) approximation in Schwarzschild spacetime 
with Schwarzschild radius $r_g = 0.3 R_0$ 
where $R_0 \sim 10^6$[cm] is the neutron star radius. 
%~\footnote{
%In the ordinal state of magnetar magnetospheres, 
%the plasma density may be too low for the MHD approximation. 
%However, 
%as mentioned below, 
%we consider an evolution of a plasma just after the onset of a giant flare of a magnetar. 
%In this case, 
%the plasma might be filled with a lot of photons and electron-positron pairs, 
%and the plasma reaches a MHD equilibrium state 
%through scatterings between the photons and pairs, 
%\textcolor{blue}{
%similarly to well coupled weakly ionized plasma, 
%in which one fluid MHD approximation is valid through the collisional coupling 
%between neutral and charged particles.
%}
%similarly to the ionized plasma of the stellar evolution, 
%which reaches its MHD state through collisions of ions and neutral particles. 
%}
In particular, 
we focus on the temporal evolution of the plasma along a closed magnetic field line. 
We set up one-dimensional closed magnetic flux tubes (Figure \ref{fig:2.1.1}) 
whose structure is fixed during our simulations
\footnote{
In this paper, 
we focused on the study of the effects of MHD waves on the dynamical evolution of magnetar magnetospheres, 
so that we assumed the  pure poloidal flux tubes. 
Extension to a background flux tube with twist is our future work.
%In this paper, 
%we assume a pure poloidal flux tube to focus on the effects of waves, 
%so that twisted components are expressed only through Alfv\'en waves. 
%Recently, 
%the twisted magnetic flux rope model (refs) is considered to be one of the most promising candidate for magnetar flare phenomena 
%because the twisted component allows to trigger various kinds of plasma instabilities (e.g. \citet{2004A&A...413L..27T,2006PhRvL..96y5002K}). 
%In this paper, 
%we focused on the study of the effects of MHD waves on the dynamical evolution of magnetar magnetospheres, 
%so that we assumed a pure poloidal flux tube. 
%Extension to the twisted background flux tube is our future work. 
}. 
The simulation region is from the surface ($r = R_0$) to the loop top that is located at $r = R_{\mathrm{top}}$ 
where $R_{\mathrm{top}}$ is the radial coordinate of the loop top. 
We assume the mirror symmetry across the loop top 
and prescribe the reflection boundary condition there. 
From the surface, 
we inject transverse velocity perturbations 
which excite Alfv\'en waves. %, as explained later in this subsection. 
To simplify the numerical modeling, 
we assume that the gravitational force is parallel to the flux tube. 
Although this approximation is not very good near the loop top, 
the gravitational force there is considerably weaker than in the near-surface region. 
Therefore, 
this approximation does not affect so much the overall trends of the evolution of the plasma in the magnetosphere. 

In order to model general configurations of closed magnetic loops, 
we introduce a super-radial expansion factor, $f(r)$, 
which was originally adopted to treat open magnetic flux tubes on the Sun \citep{1976SoPh...49...43K}. 
%\footnote{
%\textcolor{blue}{
%In this paper, we consider the temporal evolution of a magnetar magnetosphere after a giant flare. 
%In this case, 
%the neutron star surface is expected to be a gas state by the energy of the giant flare 
%and the magnetic field structure is different from the ordinal dipole field. 
%Unfortunately, 
%there is no observational data indicating the magnetic field structure during the giant flare. 
%Hence, we use the magnetic field structure similar to the solar magnetosphere 
%since the solar magnetosphere is the unique observable one. 
%}}. 
The solar surface is occupied by many small-scale closed loops, 
and as a result, open magnetic flux tubes rapidly expand around the height of these small loops, 
which is actually observed on the Sun~\citep{2008ApJ...688.1374T,2010ApJ...719..131I}. 
We expect that the large-scale magnetic loops, 
which we are now considering, would have similar structure. 
We adopt the same functional form,
%'±'±'ÉŒ»Ý'̘_•¶'ÌŽ®(2),
\begin{equation}
  \label{eq:2}
  f(r) = \frac{f_{\mathrm{Max}} \exp \left[(r - R_1) / \Sigma \right] + f_1}{\exp \left[(r - R_1) / \Sigma \right] + 1}
  ,
\end{equation}
where $f_{\mathrm{Max}}$ is the total expansion factor, and
%%'±'±'ÉŒ»Ý'̘_•¶'ÌŽ®(3).
%\begin{equation}
%  \label{eq:3}
%  f_1 = 1 - (f_{\mathrm{Max}} - 1) \exp \left[ - \frac{R_1 - R_0}{\Sigma} \right]
%  , 
%\end{equation}
the value of $f_1$ is determined so as to satisfy $f(R_0) = 1$
. 
$f(r)$ changes its value mainly at $r = R_1$, 
within the region between $R_1 - \Sigma$ and $R_1 + \Sigma$. 
We perform simulations in the super-radial expanding flux tubes with $f_{\mathrm{Max}}=7.5$ 
and in the purely radial expanding case ($f_{\mathrm{Max}}=1$). 
Throughout the paper, 
we adopt $\Sigma = 0.05 R_0$ and $R_1 = 1.1 R_0$ 
that is of the same order of the scale height. 
Then, from the conservation of radial magnetic flux, 
the radial magnetic field strength can be fixed as
%'±'±'ÉŒ»Ý'̘_•¶'ÌŽ®(1),
\begin{equation}
  \label{eq:1}
%  B_r r^2 \sqrt{1 - r / r_g} f(r) = \mathrm{const.}
  B^r r^2 f(r) = \mathrm{const.}
  ,
\end{equation}
where $r$ is the coordinate measured along the flux tube.

We divide the numerical domain into homogeneous numerical meshes with
$\Delta_r = 10^{-3} R_0$.
Regarding the inner boundary, 
mass inflow into the simulation domain is allowed, 
%which resulted from evaporation of the star surface due to large flare phenomena; 
%which results from evaporation of the star surface due to some burst phenomena in the interior of the neutron star \textcolor{blue}{that triggers giant flares}; 
which results from evaporation of the star surface due to a starquake that triggers giant flares; 
%such as a starquake of the neutron star and magnetic reconnection}; 
in addition, 
we consider transverse velocity fluctuations excited by random motions on the star surface with a power law spectrum 
indicated in the solar activity~\citep{2010ApJ...716L..19M}, $P(\nu) \propto \nu^{-1}$, 
where $\nu$ is the oscillation frequency, 
which produces linearly polarized Alfv\'en waves in the atmosphere and magnetosphere region. 
Basically, we use the average velocity, $\langle \delta v_{\perp} \rangle / c_{s,r} = 0.1, 0.5, 0.75, 0.9$ %, is listed on the table \ref{list:1}. 
where $c_{s,r} = c / \sqrt{3}$ is the relativistic sound velocity. 
We resolve an Alfv\'en wave by at least $100$ grids per wavelength 
whose frequency is approximately $10 c_A / R_0$ 
where $c_A$ is the Alfv\'en velocity. 
We discuss the validity of the assumed Alfv\'en wave properties in Section \ref{sec:sec4.3}. 
%This is equivalent to considering fluctuations on the star surface traveling at a velocity $0.5 c_s$ 
%in the region within $0.03 R_0$. 

In this paper, 
%we consider a relativistically hot star surface heated by some flare processes, 
%such as the giant flare of the magnetar. 
%we consider a relativistically hot magnetar surface heated by some burst phenomena. %a giant flare. 
we consider a relativistically hot magnetar surface heated by some burst phenomena, such as a starquake. %a giant flare. 
%For an atmosphere at the surface, , 
%we assumed its temperature is $T = 50$ [keV], 
%which is equivalent to the non-thermal component of the observed X-ray spectrum of SGRs; 
%%and we consider this is heated ; 
%for a plasma just outside of the atmosphere, 
%we assumed its temperature is $T = 5$ [keV], 
%which is equivalent to the blackbody component of the observed X-ray spectrum. 
%\textcolor{blue}{
%At the initial time, 
%we assume 
%there is an atmosphere near the surface 
%that is heated by the central star surface with a fixed temperature $T = 50$ [keV], 
%which is equivalent to the non-thermal component of the observed X-ray spectrum of SGRs; 
%for a plasma just outside of the atmosphere, 
%we assume its temperature is $T = 5$ [keV], 
%which is equivalent to the blackbody component of the observed X-ray spectrum. 
%Note that 
%the temperature of the plasma outside of the atmosphere is heated by various waves and shocks 
%and changes its value as shown in the numerical results. 
%}
At the initial time, 
we assume 
there are three regions: the star surface, the atmosphere and the plasma in the magnetosphere, 
%as shown in the left top panel of the Figure \ref{fig:3.1.1}. 
as shown in the Figure \ref{fig:2.1.1}. 
At $r = R_0$, 
we consider the star surface as the inner boundary explained above. 
In the region close to the star surface, $R_0 < r < 1.005 R_0$
~\footnote{In this case, 
the scale height of the atmosphere $\Delta R / R_0 \sim (k_B T / m c^2) (R_0 / r_g)$ is about $0.6$. 
We, however, use a conservative value $\Delta R / R_0 = 0.01$. 
We consider this initial condition does not cause any problems 
because the above initial atmosphere is blown up by the strong pressure gradient force 
immediately after a few initial timesteps as shown in the next section.}
, 
%%we assume there is an atmosphere of the central star  
%%that is heated by the central star surface with a fixed temperature $T = 50$ [keV]
%%~\textcolor{blue}{(R C.Duncan 2004,Mereghetti 2008, N. Rea \& P. Esposito 2011 and references therein)}, 
%%which is equivalent to the non-thermal component of the observed X-ray spectrum of magnetars. 
%we assume there is an atmosphere of the central star with an initial temperature $T = 50$ [keV]
%~\citep{2004cetd.conf..285D,2008A&ARv..15..225M,2011heep.conf..247R}, 
%%(R C.Duncan 2004,Mereghetti 2008, N. Rea \& P. Esposito 2011 and references therein), 
%%which is equivalent to the non-thermal component of the observed X-ray spectrum of magnetars. 
we assume there is an atmosphere heated by the central star surface with a fixed temperature $T = 50$ [keV] 
which is observed in the case of the initial spike of giant flares~\citep{2006csxs.book..547W,2008A&ARv..15..225M}. 
\footnote{
Here we assume that 
the star surface always has the temperature, $T = 50$ [keV], during timescale of our calculations. 
The energy source of the temperature can be some burst phenomena, such as a starquake. % in the neutron star core or crust 
~\citep{2004cetd.conf..285D}. 
}
In the region outside of the atmosphere, $1.005 R_0 < r < R_{\mathrm{top}}$, 
we consider the plasma in the magnetosphere with temperature is $T = 7.5$ [keV] 
%which is equivalent to the blackbody component of the observed X-ray spectrum. 
which is equivalent to the persistent blackbody component of the observed X-ray spectrum of magnetars. 
%of the soft X-ray pulses of the anomolous X-ray pulsar (AXP)  
Note that 
the temperature of the plasma outside of the atmosphere is heated by various waves and shocks 
and changes its value as shown in the numerical results. 
The initial structure of the plasma outside of the atmosphere is determined using hydrostatic equilibrium of an isothermal plasma. 
%Here we assume that 
%a burst phenomenon is triggered in the neutron star core or crust 
%and the plasma in the magnetosphere is heated mainly at the star surface. 
%The initial structure of the plasma outside of the atmosphere is determined using hydrostatic equilibrium of an isothermal plasma. 
Concerning the magnetization parameter $\sigma \equiv B^2 / 4 \pi \rho h \gamma^2 c^2 v$ %F_{\mathrm{Poynting}} / F_{\mathrm{particle}}$ 
%where $F_{\mathrm{Poynting}}$ is the Poynting energy flux and $F_{\mathrm{particle}}$ is the particle energy flux, 
where $\rho$ is the rest mass density, $h$ is the specific enthalpy, and $\gamma$ is the Lorentz factor,
we consider a moderately magnetized plasma, $\sigma = 2$ at the inner edge of the high temperature atmosphere, 
and $\sigma = 10$ at just outside of the atmosphere. 
Even though these values are too small for the magnetar magnetosphere~\citep{1995MNRAS.275..255T}, 
we believe that our findings are still be applicable to the magnetar flare phenomena. 
We will discuss it in Section 4. 

\begin{figure*}[top]
 \centering

  \includegraphics[width=6.5cm,clip]{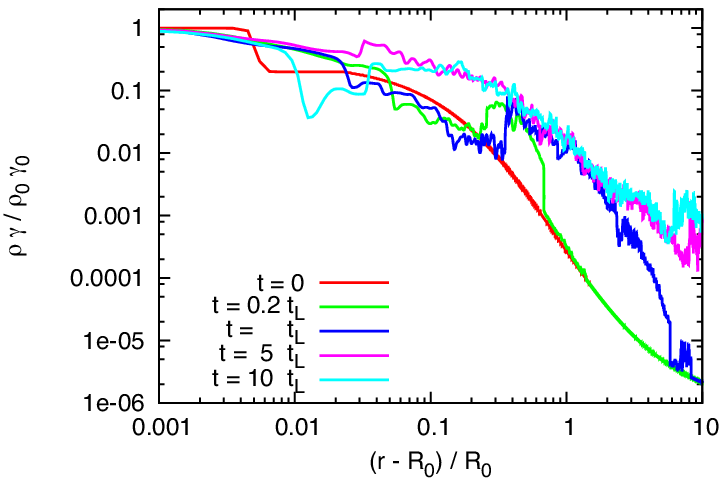}
  \includegraphics[width=6.5cm,clip]{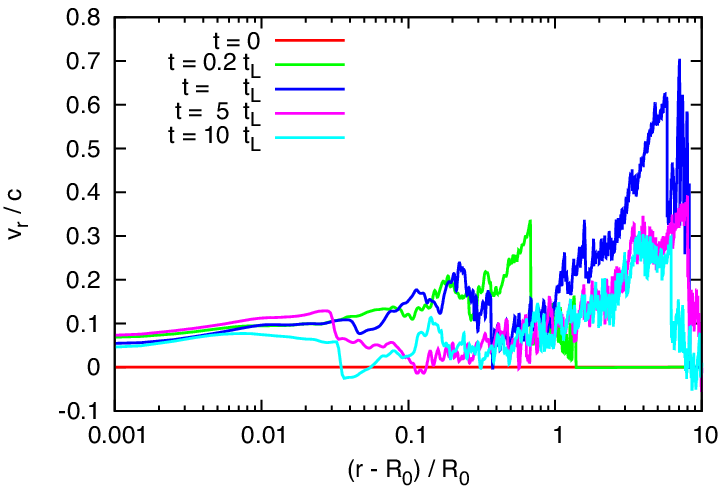}
  \includegraphics[width=6.5cm,clip]{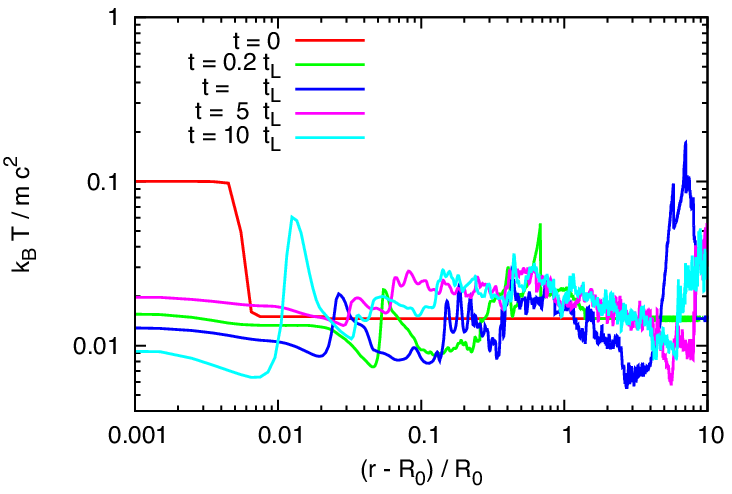}
  \caption{The snapshots of the temperature profile $k_B T / m c^2$, density profile in the laboratory frame $\rho \gamma$, and radial velocity $v_r$ 
           at $t / t_{L} = 0, 0.2, 1, 5, 10$ where $t_{L} = 10 t_0 \equiv 10 R_0 / c$ is the light crossing time along the loop. 
          }
  \label{fig:3.1.1}
\end{figure*}

\subsection{\label{sec:sec2.2}Basic Equation}

As we mentioned above, 
we solve the dynamical evolution of the plasma and waves by solving the ideal GRMHD equations 
in Schwarzschild spacetime: 
\begin{eqnarray}
  \label{eq:2.2.0}
  ds^2 &=& g_{\mu \nu} d x^{\mu} dx^{\nu} 
  , 
  \\
  &=& - w(r) dt^2 + w(r)^{-1} dr^2 + r^2 (d \theta^2 + \sin^2 \theta d \phi^2)
  ,
  \nonumber
\end{eqnarray}
where $w(r) = 1 - r_g/r$, 
$r_g \equiv 2 M G / c^2$ is the Schwarzschild radius, 
$M, G, c$ are the central star mass, the gravitational constant and the light velocity, respectively. 
We use the metric signature, $(-,+,+,+)$, along with units $c = 1$. 

The basic equations are the law of the mass density conservation 
\begin{equation}
  \label{eq:2.2.1}
   \nabla_{\mu} N^{\mu} = \frac{1}{\sqrt{- g}} \partial_{\mu} [\sqrt{- g} N^{\mu}] = 0
   ,
\end{equation}
where $N^{\mu} \equiv \rho u^{\mu}$ is the mass flux vector, 
$\rho$ is the mass density in the fluid comoving frame, 
$u^{\mu}$ is the 4-velocity 
which satisfies $u^{\mu} u_{\mu} = - 1$; 
the law of the energy and momentum conservation 
\begin{equation}
  \label{eq:2.2.2}
   \nabla_{\mu} T^{\mu}_{\nu} = \frac{1}{\sqrt{- g}} \partial_{\mu} [\sqrt{-g} T^{\mu}_{\nu}] - \frac{1}{2} \partial_{\nu} g_{\mu \rho} T^{\mu \rho} = 0
   ,
\end{equation}
where $\nabla_{\mu}$ is the covariant derivative and $T^{\mu \nu}$ is the energy-momentum tensor defined as 
\begin{equation}
  \label{eq:2.2.3}
   T^{\mu \nu} = (\rho h  + |b|^2) u^{\mu} u^{\nu} + (p_{\mathrm{gas}} + |b^2| / 2) g^{\mu \nu} - b^{\mu} b^{\nu}  
   ,
\end{equation}
where $h$ is the specific enthalpy, 
$p_{\mathrm{gas}}$ is the gas pressure, 
$\sqrt{4 \pi} b^{\mu} = (^*F^{\mu \nu}) u_{\nu}$ is the magnetic field 4-vector, 
$F^{\mu \nu}$ is the electromagnetic field tensor, 
$^*F^{\mu \nu}$ is the dual of the electromagnetic field strength tensor 
given by $^*F^{\mu \nu} = 1/2 \epsilon^{\mu \nu \gamma \delta} F_{\gamma \delta} = \sqrt{4 \pi} (u^{\mu} b^{\nu} - u^{\nu} b^{\mu})$, 
and $|b^2| = b^{\mu} b_{\mu}$. 
The induction equation is
\begin{equation}
  \label{eq:2.2.4}
  \nabla_{\mu} (^*F^{\mu \nu}) = \frac{1}{\sqrt{- g}} \partial_{\mu} [\sqrt{-g} (^*F^{\mu \nu})] = 0
  .
\end{equation}

We use the relativistic HLLD scheme~\citep{2009MNRAS.393.1141M} %\footnote{
%Of cause, 
%this HLLD flux is not a very good approximation for the numerical flux 
%comparing with the case of the flat geometry. 
%However, we expect this gives us a better estimate of the numerical flux. 
%For this purpose, 
%we introduce pseudo velocity 4-vector $U^{\mu}$ 
%which is related to the 4-velocity vector $u^{\mu}$ as $U^{\mu} = (\sqrt{w} u^t, u^r / \sqrt{w}, r u^{\theta}, r \sin \theta u^{\phi})$. 
%This new vector satisfies the following relation: $- (U^t)^2 + (U^r)^2 + (U^{\theta})^2 + (U^{\phi})^2 = -1$,  
%and can play a role as the velocity 4-vector necessary for the calculation of the HLLD flux. 
%Note that this does not mean we use the FIDOs frame 
%since we does not change spacetime coordinate $(t,r,\theta, \phi)$ but just introduce a vector for convinience of the calculation of the numerical flux. 
%We also use a similar vector for the magnetic field vector. 
%} 
to calculate approximated values of the numerical flux of the conservative part 
which is multiplied by $\sqrt{-g}$ in Equation (\ref{eq:2.2.1}), (\ref{eq:2.2.2}), (\ref{eq:2.2.4}). 
The detailed numerical method is presented in Appendix A. %\ref{sec:secA.1}. 
For the equation of state, 
we assume the relativistic ideal gas with $h = 1 + (\Gamma / (\Gamma - 1))(p_{\mathrm{gas}} / \rho)$ 
where $\Gamma = 4 / 3$. 
%Since we assume a magnetosphere with a dense plasma, 
%we also assume the relativistic adiabatic heat ratio. 

\section{\label{sec:sec3}Results}

In this section, 
we present results of our numerical simulations. 
In Section \ref{sec:sec3.1}, 
we present our numerical results in the case of $R_{\mathrm{top}} = 11 R_0$ 
where the temperature at the loop top shows the maximum value. 
In Section \ref{sec:sec3.2}, 
we present the temporal evolution of the temperature at the loop top. 

%In addition, 
%we give an implication in the giant flare of SGRs. 
\subsection{\label{sec:sec3.1}Temporal Evolution of Physical Variables in the Magnetosphere}

Each panel of Figure \ref{fig:3.1.1} is the temporal evolution of the temperature $k_B T / m c^2$, 
density in the neutron star frame $\rho \gamma / \rho_0 \gamma_0$, 
and radial velocity $v_r / c$ %and plasma $\beta \equiv p_{gas} / (b^2 / 8 \pi)$ 
%at $t / t_L = 0, 0.2, 1, 5, 10$ in the case of $\delta v_{\perp} = 0.75 c / \sqrt{3}$ %of run CS075G20 
at $t / t_L = 0, 0.2, 1, 5, 10$ in the case of $\langle \delta v_{\perp} \rangle = 0.75 c_{s,r}$ %of run CS075G20 
where $\rho_0, \gamma_0$ are the initial rest mass density and Lorentz factor at the surface, respectively, 
%$b^2 = b_{\mu} b^{\mu}$ is the square of the 4-magnetic field vector explained in Section \ref{sec:sec2.2}, 
and $t_{L} = 10 t_0 \equiv 10 R_0 / c$ is the light crossing time of the flux tube. 
%The figures show that 
%after starting the calculation 
In this calculation, 
Alfv\'en waves driven at the surface propagate along the flux tube nearly at the light velocity  
and the flux tube is gradually filled with the waves. 
As propagating outwards, 
the amplitude of the Alfv\'en waves grows due to the decrease of the background density. 
%When the amplitude of the Alfv\'en waves reached to a finite size at which non-linear effects become important, 
%they decay into Alfv\'en waves and slow waves (Sagdeev and Galeev 1969, Goldstein 1978, Terasawa and Hoshino 1986, Suzuki and Inutsuka 2005, 2006); 
When the normalized amplitude, $\langle \delta v \rangle / c_A$, of the Alfv\'en waves increases to $\gtrsim 0.1$, 
nonlinear effects become gradually important, 
e.g., they decay into fast and slow waves and counter-propagating Alfv\'en waves 
due to their magnetic pressure 
\citep{1969npt..book.....S,1978ApJ...219..700G,1986JGR....91.4171T}; 
as is reported in the work studying the solar corona and solar wind. 
We find that 
these fast and slow waves evolve into shocks and convert their wave energy into the thermal energy through the shock dissipation 
\citep{1999ApJ...514..493K,2004ApJ...601L.107M,2005ApJ...632L..49S,2006JGRA..111.6101S,2008ApJ...678.1200S}. 
%The Alfv\'en waves also transfer matters from the central star surface to the loop top. 
The Alfv\'en waves also play a role in transferring matter from the central star surface to the loop top 
by the associated magnetic pressure (ponderomotive force). 
\begin{figure}[h]
 \centering
  \includegraphics[width=9cm,clip]{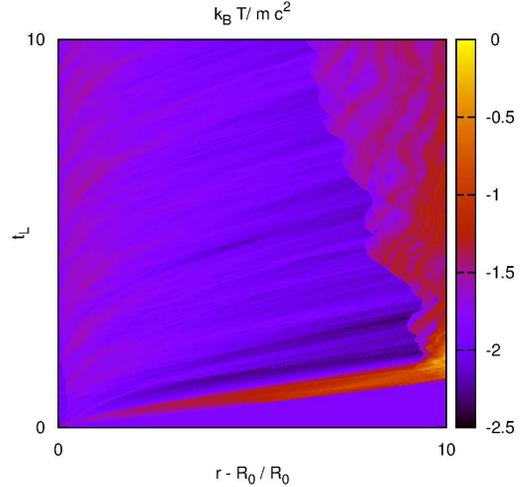}
  \caption{The time evolution of the temperature profile $k_B T / m c^2$ in the case of $R_{\mathrm{top}} = 11 R_0$. 
           The horizontal axis is the radius $(r - R_0)/ R_0$, 
           and the vertical axis is time measured in the light crossing time $t_L \equiv 10 t_0$. 
           The left-hand edge is the neutron star surface and the right-hand edge is the top of the flux tube. 
           The red region emerging at the origin is the region between the front edge of the Alfv\'en waves 
           and the initially generated slow shock front.            
          }
  \label{fig:3.1.2}
\end{figure}
Figure \ref{fig:3.1.2} is the time evolution of temperature profile. 
The red belt in the lower region of this figure is a hot region heated by the Alfv\'en waves. 
The lower boundary of this region corresponds to the front edge of Alfv\'en waves 
and the upper boundary to the slow shock front generated on the neutron star surface at the initial time. 
This figure shows that 
the highest temperature resulted from the collision of the hot regions at the loop top. 
This temperature rising continues until the arriving of the slow shock, 
and afterwards this region gradually expands absorbing the subsequent Alfv\'en waves. 

As is well known, 
waves have their own energy and momentum flux 
and can produce an extra pressure called '\textit{wave pressure}' 
whose typical form is given by $P_w = \rho h \langle \delta v^2 \rangle / 2$. 
Following the traditional wave driven wind theory~\citep{1999isw..book.....L}, 
as waves propagate outwards in stellar atmospheres with decreasing density, 
the wave pressure gradient is generally generated. 
%as waves propagate outwards in steller atomospheres 
%the density gradient in the atmosphere results in the wave pressure gradient 
This wave pressure gradient induces a force on the background plasma 
and this results in a transonic outgoing flow called '\textit{wave driven wind}'. 
Differently from the ordinary wave driven wind cases, 
we impose the reflection boundary at the loop top, 
so that the radial velocity is not transonic. 
The radial velocity profile, however, shows that 
the Alfv\'en waves produce the wave-pressure and drive upgoing flows. 
Note that 
these upgoing flows play an essential role in the mass supply to the loop top regions, 
which triggers giant flare-like phenomena as discussed below. 

In addition to the Alfv\'en waves, 
we find that 
a slow shock induced by the initial high-temperature atmosphere plays a very important role 
for the evolution of the magnetosphere. 
Because of the initial pressure gradient between the atmosphere and the magnetosphere plasma, 
a slow shock is generated just after starting the calculation. 
This slow shock propagates outwards approximately about the relativistic sound velocity $c_{s,r} \equiv c / \sqrt{3}$, 
%which is equivalent to the sound velocity with the temperature $0.1 k_B T / m c^2$,  
and heats the plasma 
(The slow shock is at $r - R_0 \sim 0.6, 7$ when $t = 0.2 t_L, t_L$ in Figure \ref{fig:3.1.1}). 
When this slow shock reaches the loop top, 
the background plasma is heated drastically by the collision of two slow shocks\footnote{ 
Note that plasma evolved in the same manner along the other side of the flux tube due to the reflection boundary condition at the loop top.}, 
which results in the initial spike of flares as is explained later; 
the plasma temperature around the loop top becomes more than $0.1 m c^2$. % and the plasma $\beta$ became larger than $1$. 
After the shock reaches the loop top, 
the plasma reaches a quasi-steady state 
and the plasma continues to be heated slowly by slow shocks generated by Alfv\'en waves
\footnote{
In relativistic Poynting-dominated plasma, 
the fast-wave velocity becomes close to the light velocity, 
so that fast shocks are very weak in this case and does not influence on the background plasma at all. 
On the other hand, 
the slow wave velocity is at most the relativistic sound velocity even in Poynting-dominated plasma. 
This means slow shocks can be very strong even in Poynting-dominated plasma, 
and can convert almost all energy of the magnetic field perpendicular to their shock front.
}
. 
\subsection{\label{sec:sec3.2}Temperature Evolution at the Loop Top}

%\begin{figure*}[top]
% \centering
%  \includegraphics[width=3.5cm,clip]{./graphics/wave_T.eps}
%  \includegraphics[width=5cm,clip]{./graphics/wave_T_amp075_10R0.eps}
%  \includegraphics[width=7.5cm,clip]{./graphics/wave_T_amp075_20R0.eps}
%  \caption{The snapshots of the temperature profile $k_B T / m c^2$, density profile $\rho \gamma$, radial velocity $v_r$ and plasma $\beta$ 
%           at $t / t_0 = 0, 0.2, 1, 5, 20$. 
%          }
%  \label{fig:3.1.1.2}
%\end{figure*}

\begin{figure}[h]
 \centering
  \includegraphics[width=8cm,clip]{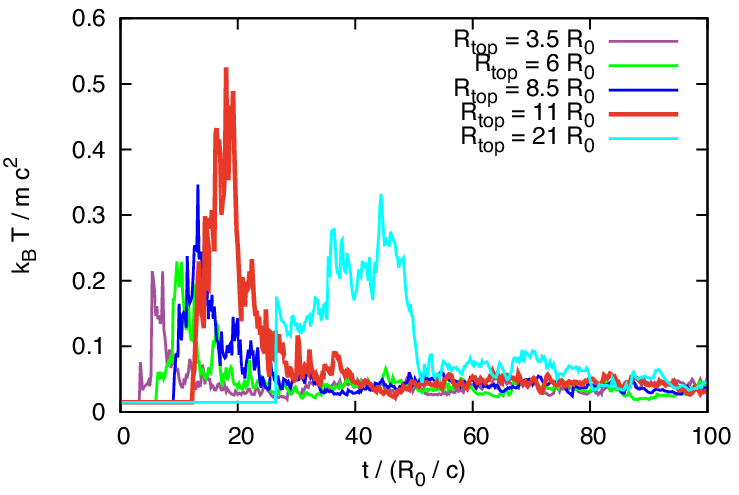}
  \includegraphics[width=8cm,clip]{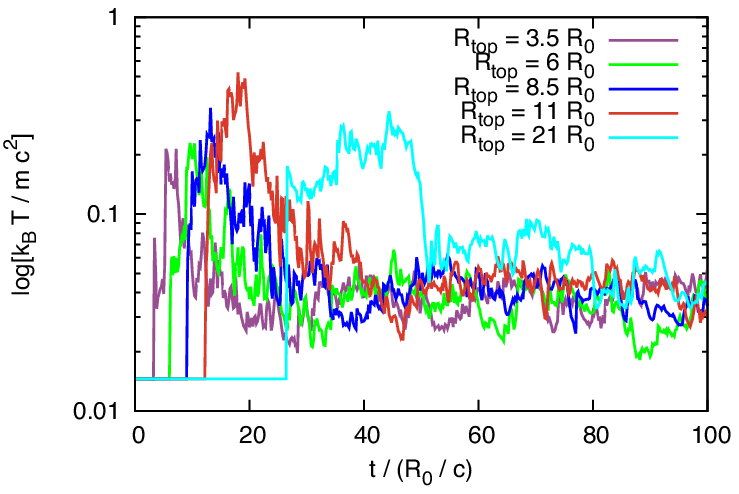}
  \caption{The temporal evolution of the temperature $k_B T / m c^2$ at the loop top: 
%           using different length flux tube. 
%           Top panel: $R_{top} = 11 R_0$. 
%           Bottom panel: $R_{top} = 3.5, 6, 8.5, 11, 21 R_0$. 
           $R_{\mathrm{top}} = 3.5, 6, 8.5, 11, 21 R_0$. 
           We assumed $\langle \delta v_{\perp} \rangle = 0.75 c_{s,r}$. 
           Top: normal-scale. Bottom: log-scale
          }
  \label{fig:4.1.2}
\end{figure}

In this paper, 
we assumed the existence of the initial high temperature atmosphere %and plasma in the magnetosphere 
%resulted from the SGR fireball. 
%resulting from a SGR flare phenomenon 
%which triggers a giant flare. 
resulting from some burst phenomena. 
As explained in the previous section, 
this structure induces a strong slow shock 
and this slow shock heats plasma in the magnetosphere. 
In this section, 
we focus on the evolution of the temperature at the loop top 
and study its implication in the giant flare. 
Here we assume the existence of a mechanism triggering the giant flare 
and we do not discuss its origin. 

Figure \ref{fig:4.1.2} is the temporal evolution of the loop top temperature 
using various loop size: $R_{\mathrm{top}} = 3.5, 6, 8.5, 11, 21 R_0$. 
%The top panel of Figure. \ref{fig:4.1.2} is the temporal evolution of the temperature at the loop top 
%which is at $R_{top} = 11 R_0$ in the case of $\delta v_{\perp} = 0.75 c / \sqrt{3}$. 
First we note that 
the temperature keeps its initial value %$k_B T / m c^2 = 0.015$ 
%until $t \sim (R_{top} - R_0) / c_A = 10 R_0 / c$, or the Alfven crossing time along the loop. 
until $t \sim (R_{\mathrm{top}} - R_0) / c_A$, or the Alfv\'en crossing time along the loop. 
Around $t = (R_{\mathrm{top}} - R_0) / c_A$, 
the Alfv\'en waves emitted at initial time reach the loop top 
and collides with waves 
that propagate from the other side of the loop as indicated in Figure \ref{fig:3.1.2}. 
Note that the Alfv\'en waves are accompanied by hot plasma regions 
%resulted from the dissipation of them as shown in Appendix \ref{sec:secA.2}, 
resulted from the shock heating and the Alfv\'en waves' dissipation as shown in Appendix B, %\ref{sec:secA.2}, 
the compression by the collision of this region induces the initial temperature rise. 
%Around $t \sim 10 R_0 / c$, 
The temperature reaches its maximum value of about $0.3 k_B T / m c^2$ at $t \sim (R_{\mathrm{top}} - R_0) / c_{\textrm{slow}}$ 
%The temperature reaches its maximum value about $0.5 k_B T / m c^2$ at $t \sim (R_{top} - R_0) / c_s \sim 20 t_0$ 
%This collision of two slow shocks induces a sharp temperature rising up to about $1.6 m c^2$. 
%and this temperature rising lasts until $t \sim (R_{top} - R_0) / c_s \sim 15 t_0$ 
where $c_{\textrm{slow}} \sim c_{s,r} \equiv c / \sqrt{3}$ is the slow shock velocity observed in our simulation, 
which is equivalent to the time when 
the slow shock induced by the initial pressure discontinuity at the atmosphere 
reaches the loop top. %and collides to another slow shock 
Afterwards, 
the temperature shows quasi-stable behavior with small perturbations by the Alfv\'en waves. 

This panel shows that 
the maximum temperature increases with loop size until $R_{\mathrm{top}} = 11 R_0$; 
in the case of $R_{\mathrm{top}} = 21 R_0$, 
the maximum temperature is smaller than that of $R_{\mathrm{top}} = 11 R_0$, 
but the duration of the temperature rising becomes longer. 
This can be explained as follows. 
\begin{figure}[t]
 \centering
  \includegraphics[width=8cm,clip]{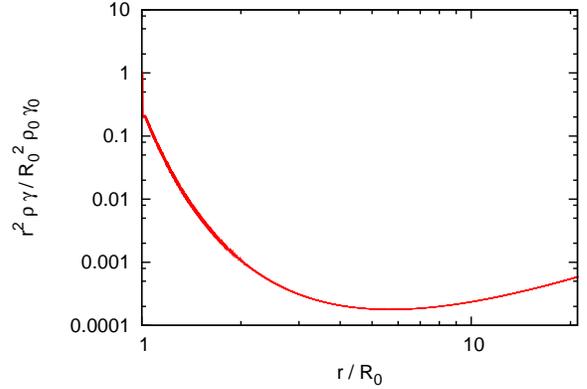}
%  \caption{A snapshot of the initial density profile per unit radial coordinate in the spherical shell $\rho r^2$. }
  \caption{A snapshot of the initial mass profile in the spherical shell per unit radial coordinate. 
          }
  \label{fig:4.1.5}
\end{figure}
Assuming the initially triggered slow shock as a blast wave, 
the conservation law of the wave energy flux can be written as, $\rho (\delta v)^2 c_s f(r) \sim \mathrm{Const.}$, 
when background velocity is sufficiently smaller than background sound velocity~\citep{1999isw..book.....L}. 
Thus, the temperature of the blast wave is 
\begin{equation}
  \label{eq:4.0}
  \delta T \sim (\delta v)^2 \propto (\rho f(r) c_s)^{-1} 
  .
\end{equation}
Note that now the wave group velocity $c_s$ takes a constant value, $c / \sqrt{3}$, 
so that the temperature of the slow shock region is in proportion to $(\rho f(r))^{-1}$. 
Figure \ref{fig:4.1.5} is the initial profile of the mass in the spherical shell per unit radial coordinate, $\rho f(r)$, 
using an approximation: $f(r) \sim r^2$. 
It shows that 
the mass profile has a minimum point around $r \sim 6 R_0$. 
Equation (\ref{eq:4.0}) indicates that 
the temperature at the loop top increases with increasing the length of the loop until it reaches the above minimum point, 
and this explains the tendency of our numerical results very well. 
%Before this minimum point, 
%the shock can keep its energy due to the energy supply from the Alfv\'en waves indicated in Appendix \ref{sec:secA.2} 
%and the decrease of the cold matter in front of the shock fronts; 
%on the other hand, 
%beyond the minimum point, 
%the amount of the cold matter in the spherical shell increases 
%and this gradually reduces the shock strength. 
Figure \ref{fig:4.1.5} shows that 
%the \textcolor{blue}{mass} change around the minimum point is not so steep, 
the radial gradient of the mass around the minimum point is not so steep, 
%so that the shock strength still increases with the loop length beyond the minimum point 
and we consider that 
this is the reason why the shock strength still increases beyond the minimum point 
as shown in the bottom panel of Figure \ref{fig:4.1.2}. 
%In the case of $R_{top} = 21 R_0$, 
%the energy decay of the shocks by the increase of cold matter finally 
%becomes larger than the energy supply by the Alfv\'en waves, 
%and the maximum temperature starts to decrease. 
Note that 
Equation (\ref{eq:4.0}) can also be applied to the Alfv\'en waves, 
so that heating by Alfv\'en waves does not change the above argument essentially. 

The above radial coordinate at the minimum mass point can be obtained as follows. 
First, if we assume the non-relativistic hydrostatic equilibrium, 
the mass profile can be written as: $\rho = \rho_0 \exp[- \frac{r - R_0}{H} \frac{R_0}{r}]$ 
where $H = (R_0 / r_g) (c_s / c)^2 R_0$ is the scale height, $\rho_0$ is the density at $r = R_0$ and $c_s$ is the sound velocity. 
Using this expression, 
the minimum point can be calculated as: 
\begin{equation}
  \label{eq:4.1}
  \frac{d}{dr} [r^2 \rho] \propto 2 \rho \left[ r - \frac{R_0^2}{2 H} \right]
  . 
\end{equation}
From this equation, 
we can obtain the radial coordinate of the minimum mass point as 
\begin{eqnarray}
  \label{eq:4.2}
  r_{\mathrm{min}} / R_0 &=& R_0 / 2 H
  \nonumber
  \\
  &\sim& 8 \left( \frac{r_g}{0.3 R_0} \right) \left( \frac{c_s(T = 7.5 \textrm{[keV]})}{c} \right)^{-2}
  . 
\end{eqnarray}
%If we substitute $k_B T / m c^2 = 0.015$, 
%If we substitute $k_B T = 7.5$ [keV], 
%this coordinate becomes $r_{min} \sim 8 R_0$. 
The minimum value in Figure \ref{fig:4.1.5} seems a little smaller than this value 
because of the general relativistic effect. 
This coordinate is a critical value 
at which the strength of the shocks change, 
so that this can be used as a characteristic coordinate of this phenomena. 

\begin{figure}[t]
 \centering
  \includegraphics[width=8cm,clip]{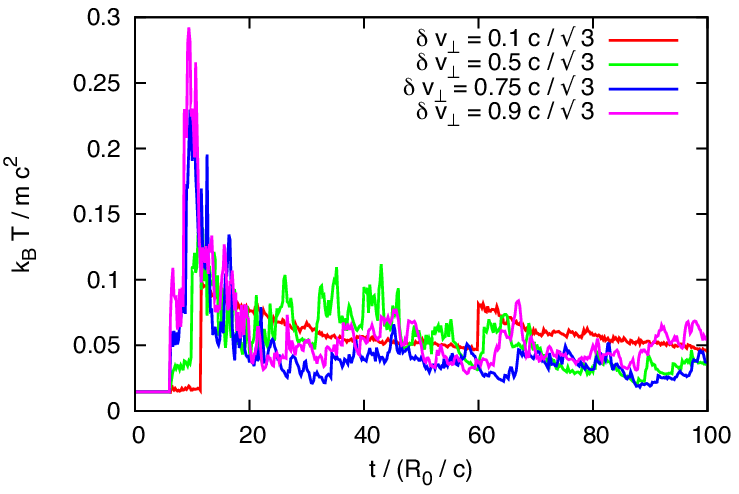}
  \includegraphics[width=8cm,clip]{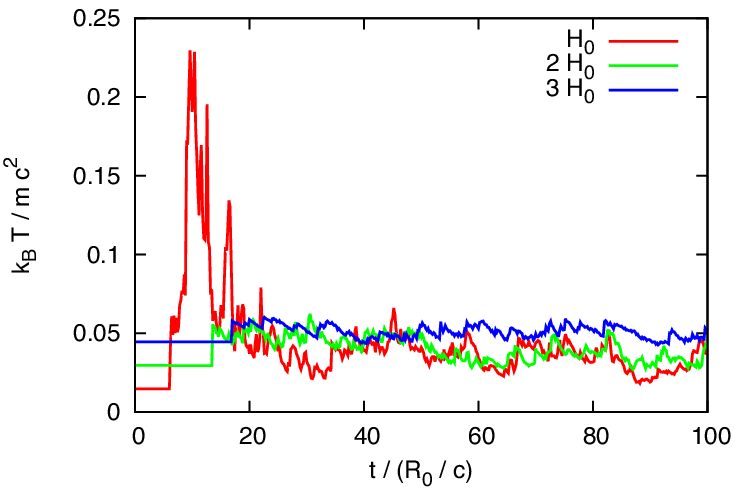}
  \caption{The temporal evolution of the temperature $k_B T / m c^2$ at the loop top $R_{\mathrm{top}} = 6 R_0$. 
           Top: amplitude dependence. 
           Bottom: scale height dependence. 
           $H_0$ is the scale height in the case of $T = 7.5$ [keV]. 
          }
  \label{fig:4.1.3}
\end{figure}

\begin{figure}[t]
 \centering
  \includegraphics[width=8cm,clip]{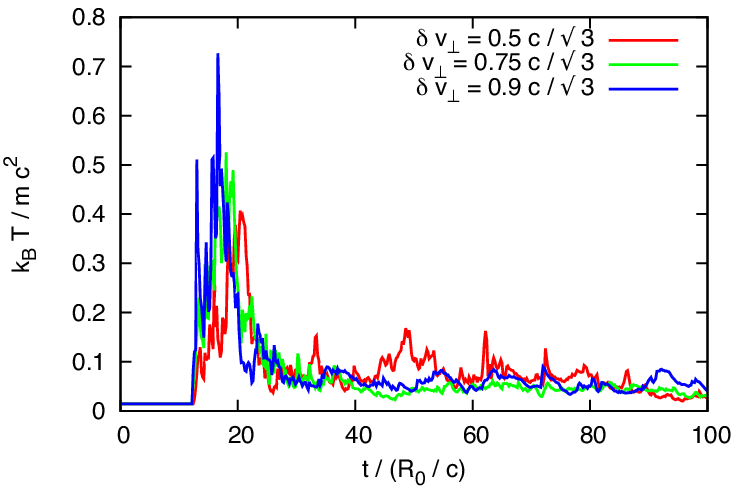}
  \includegraphics[width=8cm,clip]{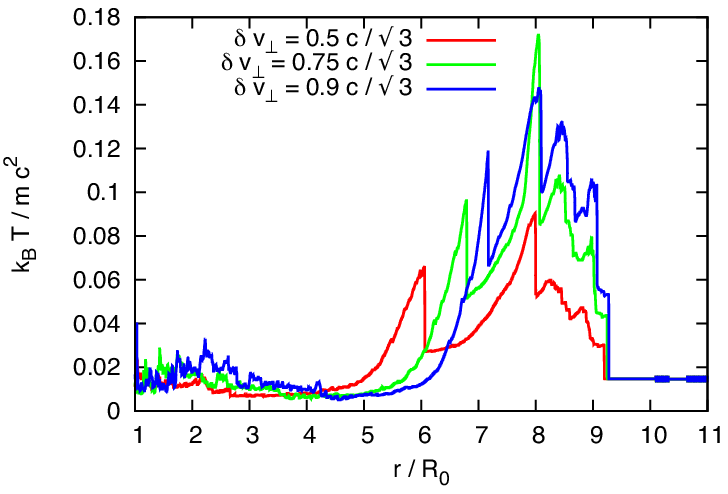}
  \caption{Top: The temporal evolution of the temperature $k_B T / m c^2$ at the loop top $R_{\mathrm{top}} = 11 R_0$
           using different wave amplitude. 
           Bottom: Snapshots of the temperature profile at $t = 10 R_0 / c$. 
          }
  \label{fig:4.1.4}
\end{figure}

Next,
we discuss their dependence on the amplitude of the Alfv\'en waves and scale height $H = (R_0 / r_g) (c_s / c)^2 R_0$. 
The top panel of the Figure \ref{fig:4.1.3} is the temporal evolution of the loop top temperature 
whose length is $R_{\mathrm{top}} = 6 R_0$ 
using different Alfv\'en wave amplitude. 
This figure shows 
the peak temperature increases with increasing the wave amplitude. 
This is because 
more energy is supplied to the hot region heated by the Alfv\'en waves 
as the amplitude of the waves increases. 
This also reduces the energy loss of the shock waves propagating through the cold plasma, 
and help to keep the strong shock condition. 
%In particular, 
On the other hand, 
when $\langle \delta v_{\perp} \rangle / c_{s,r} = 0.1$, 
the contribution from the Alfv\'en waves is significantly decreased, 
so that there is no initial spike like feature in their curve 
but steep temperature rising by the collision of the slow shocks 
that is driven by the initial pressure gradient at the atmosphere. 
%propagating radially outwards. 
Note that 
the rising time of the initial spike is moved forward as the wave amplitude increases. 
This can be explained by the same reason 
%because the peak energy increases with increasing the energy of the shocks colliding at the loop top.  
because the shock front velocity becomes faster as the temperature in a post-shock region increases. 
%\textcolor{red}{
Concerning the amplitude, 
the amplitudes used in our simulations are too large comparing with a theoretical value, $\delta v_{\perp} \sim 0.01c$, predicted by \citet{1989ApJ...343..839B}. 
However, 
even if the predicted amplitude is used, 
our mechanism works in the magnetar magnetosphere 
because of its Poynting-dominated environment 
where even Alfv\'en waves with small amplitude have much larger energy than background thermal energy; 
on the other hand, 
our simulations considered a plasma with moderate magnetization, $\sigma \leq 10$, 
so that the larger wave amplitude is necessary to produce a sufficient energy on the background plasma. 
The detailed estimate of their energetics is discussed in Section \ref{sec:sec4.3}. 
%}
%%%%%
In the bottom panel of the Figure \ref{fig:4.1.3}, 
we plot the temporal evolution of the loop top temperature using different values of $r_g / R_0$ to study the scale height dependence. 
We find the peak energy decreases with increasing the scale height. 
%and this is because the amount of cold matter in front of the shock fronts increases 
%with increasing the scale height. %relates to the energy supply from the Alfv\'en waves. 
This is because both the density and pressure in front of the shock fronts increase 
with increasing the scale height, %relates to the energy supply from the Alfv\'en waves. 
and we cannot neglect the pressure in the magnetosphere comparing with the pressure in the postshock region, 
which means the breakdown of the strong shock condition. 
%As is explained, 
%the slow shock is heated by the dissipation energy of the Alfv\'en waves. 
%The relative wave energy to the thermal energy, however, decreases 
%as the plasma $\beta$ increases. 
%The plasma $\beta$ of the magnetosphere increases with increasing the central star radius 
%because the scale height $H = c_s^2 / (G M / R_0)$ is proportional to the central star radius. 
%Note that this indicates that 
%the energy of giant flares might increase with increasing the neutron star mass. 
%the dissipation energy of the Alfv\'en waves becomes less significant comparing with the plasma thermal energy 
%and this reduces the energy supply to the slow shock effectively. 

Top panel of Figure \ref{fig:4.1.4} is the temporal evolution of the temperature at the loop top 
whose length is $R_{\mathrm{top}} = 11 R_0$. 
This panel shows that, 
%their maximum temperature shows similar dependence on the wave amplitude, 
%though the amount of their difference becomes small compared with the case of $R_{top} = 6 R_0$. 
compared with the case of $R_{\mathrm{top}} = 6 R_0$, their maximum temperature shows similar dependence on the wave amplitude, 
though the amount of their difference becomes small. 
%their total energy during the temperature rising seems increase with increasing the amplitude. 
%Differently from the loop with length $R_{loop} = 6 R_0$, 
%we cannot find any significant difference of their maximum temperature, 
%though their total energy during the temperature rising seems increase with increasing the amplitude. 
Bottom panel of Figure \ref{fig:4.1.4} is a snapshot of the temperature profile of each wave amplitude cases 
just before their collision with their counter parts from the other side of loop. 
At $r \sim 9.2 R_0$, 
there are fast shock fronts induced by the collapse of the Alfv\'en waves. 
Since the fast velocity is close to the light velocity in high-$\sigma$ plasmas, 
the fronts of the 3 cases are at the same position. 
On the other hand, 
the position of slow shock fronts, located around $r = 6 - 7 R_0$ in the figure, depends on their amplitude; 
their coordinate increases with increasing the Alfv\'en wave amplitude 
because of the energy supply to the slow shocks from the Alfv\'en waves. 
Note that in the top panel of Figure \ref{fig:4.1.4}, 
all the initial temperature rising occur at the same time. 
However, the time of their peak temperature are delayed as the wave amplitude decreases. 
We consider that 
this also supports our conclusion that 
the initial temperature rising is caused by the hot region heated by the Alfv\'en waves 
and the peak temperature is controlled by the collision of the initially induced slow shocks at the atmosphere.

\section{\label{sec:sec4}Discussion}

\subsection{\label{sec:sec4.1}Light Curves}

In Section \ref{sec:sec3.2}, 
we showed the temperature evolution at the loop top. 
%and concluded the collision of shocks at the loop top can be a candidate of the mechanisms of the initial spike of the magnetar giant flare. 
In this section, 
we discuss the possibility of the relation between the initial spike of giant flares 
and the collision of shocks at the loop top. % can be a candidate of the mechanisms of the initial spike of the magnetar giant flare. 
%Note that those curves in the bottom panel of Figure. \ref{fig:4.1.2} remember us the light curve of the giant flare. %and its spectrum. 
We note that the curves in the bottom panel of Figure \ref{fig:4.1.2} are similar to the light curve of flare phenomena. %and its spectrum. 
In addition, 
the maximum temperature of the curves reaches $100$ [keV]. 
These indicate that 
the collision of shocks at loop tops can be a candidate of the mechanism of the initial spike of magnetar giant flares. 
%In particular, 
%the sharp rising of temperature by the collision of the hot region and accompanying quasi-stable state indicate 
%some relations to the initial spike. %and pulsed tail of the giant flare. 
To examine this, %hypothesis, 
we consider the time scale of this phenomena in the case of magnetar. 
%%%
Using $R_0 \sim 10^6$[cm] and $c = 3 \times 10^{10}$[cm/s], 
the light crossing time of the central star becomes $t_0 = R_0 / c \sim 0.03$ [ms]. 
%and the rising time of the initial peak indicated in Fig. \ref{fig:4.1.1} is about $5 t_0 \sim 0.15$ [ms]. 
From the above discussion, 
%the duration of the temperature rising can be estimated as
the duration of the temperature rising, 
which corresponds to the time lag of the arrivals of the Alfv\'en waves and the slow shock, 
can be estimated as
\begin{equation}
  \label{eq:4.3}
  \Delta t \sim r_{\mathrm{min}} / c_{\mathrm{slow}} - r_{\mathrm{min}} / c_A
  .   
\end{equation}
Using the following parameters: $r_{\mathrm{min}} = 8 R_0, c_A = c, c_{\mathrm{slow}} = c_{s,r}$, 
the duration becomes $\Delta t \sim 0.3$ [ms]. 
The observed rising time of the initial peak of the giant flare is $\sim 1$ [ms], 
so the obtained value in our simulation is of the same order. %too short to explain the observed timescale. 
%However, 
Note that this conclusion is obtained using flux tubes with length $R_{\mathrm{top}} > r_{\mathrm{min}}$; 
%if we consider shorter flux tubes, 
if we take into account shorter flux tubes, 
the total duration time, which is the superposition of the flares of tubes with various length, can be longer, 
since initial rising time of shorter tubes is earlier than that of longer tubes. 
%%%

Concerning the average temperature during the collision of heated regions, 
we can estimate its value as follows. 
%%%%%%%%%%%%%%%%%%%%%%%%%%%%%%%%%%%%%%%% Alfven wave %%%%%%%%%%%%%%%%%%%%%%%%%%%%%%%%%%%%%%%%%%%%%%%%%%%%%
%First, 
%the temperature in the heated region by the Alfv\'en waves can be estimated 
%using the result in Appendix \ref{sec:secA.2}. 
%Using the expression in Appendix \ref{sec:secA.2} and the value of the magnetization parameter in the magnetosphere, $\sigma = 10$, 
%the temperature in the region is obtained as 
%\begin{equation}
%  \label{eq:4.3}
%  T_0 + \Delta T = T_0 + 0.0267 B_{\perp}^2 / 2 \rho \sim 0.04 m c^2
%  ,
%\end{equation}
%where we used $B_{\perp} = (0.75 c / \sqrt{3}) B_0$ and $k_B T_0 / m c^2 = 0.015$. 
%%%%%%%%%%%%%%%%%%%%%%%%%%%%%%%%%%%%%%%% Sedov solution %%%%%%%%%%%%%%%%%%%%%%%%%%%%%%%%%%%%%%%%%%%%%%%%%%%%%
First, 
the slow shock initially driven by the star surface is a strong shock propagating into the cold magnetosphere along the radially-expanding closed loop. 
This can be approximated by the non-relativistic strong shock. %Sedov solution~\citep{1959sdmm.book.....S}. 
The temperature just inside of slow shocks can be written as: 
\begin{equation}
  \label{eq:4.4.1}
  k_B T / m c^2 = 2 \frac{\Gamma - 1}{(\Gamma + 1)^2} \left( \frac{u_{\textrm{sh}}}{c} \right)^2 
  \simeq 0.04 \left( \frac{u_{\textrm{sh}}}{c_{s,r}} \right)^2
  ,
\end{equation}
where we used $\Gamma = 4/3$ and $u_{\textrm{sh}}$ is the shock velocity in the upstream comoving frame. 
Here we used the observed value $c_{s,r}$
%\footnote{
%The original Sedov solution also gives us the explicit expression of the shock velocity. 
%In our calculation, 
%the heating by the Alfv\'en waves changes the evolution of the shock velocity, 
%so that we used the numerical value obtained in our calculation. 
%}
. 
Note that this value approximately reproduces the result in Figure \ref{fig:3.1.1} (see around $r = 5 - 6 R_0$ of the blue line in the left-bottom panel). 
Next, 
the collision of the above region heated by the Alfv\'en waves results in two slow shock fronts 
due to their supersonic velocity 
and the post-shock region behind the shock fronts is equivalent to the temperature rising state in Figure \ref{fig:4.1.2}. 
Since the nature of the slow shock in high-$\sigma$ plasmas is close to hydrodynamic shocks, 
we derive the post-shock temperature using the results of the pure hydrodynamic Riemann problem
~\citep{1994JFM...258..317M,2005ApJS..160..199M}. 
Following the relativistic Rankine-Hugoniot relation, 
the post-shock pressure can be written as follow: 
\begin{eqnarray}
  \label{eq:4.4}
  p_{ps} &=& p_{u} + \frac{h_{ps} \gamma_{ps} v_{ps} - h_u \gamma_u v_u}{\zeta}
  ,
  \\
  \label{eq:4.5}
  \frac{1}{\rho_{ps} \gamma_{ps}} &=& \frac{1}{\rho_u \gamma_u} - \zeta (v_{ps} - v_u)
  ,
\end{eqnarray}
where subscript ``u'' and ``ps'' means the variables are in upstream region and post-shock region, respectively, 
and 
\begin{equation}
  \label{eq:4.6}
  \zeta = \frac{1}{\rho_u \gamma_u (v_{sh} - v_u)}
  . 
\end{equation}
In the above expression, 
$v_{sh}$ is the velocity of the shock front. 
Since the heated region collides with the counter part propagating from the other side of the loop, 
the flow velocity in the post-shock region is $v_{ps} = 0$ due to the conservation of momentum. 
For simplicity, we assume the shock front velocity $v_{sh} = - c_{s,r}$ 
%that is the relativistic limit value in the postshock rest frame and propagates in the inward direction. 
.
In this case, 
using the above equations, 
we can obtain the following expression of the post-shock temperature: 
\begin{equation}
  \label{eq:4.7}
  \frac{k_B T_{ps}}{m c^2} = \frac{p_{ps}}{\rho_{ps}} 
  = \frac{p_u}{\rho_u \gamma_u} \left[ 1 - \frac{v_u}{c_{s,r} + v_u} \right] + c_{s,r} h_u \gamma_u v_u
  .
\end{equation}
By substituting $T_u = 0.04 m c^2$, obtained in Equation (\ref{eq:4.4.1}), and $v_u = 2 v_{sh} /(\Gamma + 1)$ 
that is the post-shock velocity in the case of strong shocks, %Sedov value of the post-shock velocity, 
%~\footnote{
%Here we use the relativistic sound velocity as a characteristic velocity. 
%}, 
we obtain $T_{ps} \sim 0.4 m c^2$ 
that reproduces the average value of the top panel of Figure \ref{fig:4.1.2} well. 
Although this value is much smaller than that in the case of $R_{\mathrm{top}} = 11 R_0$, 
we consider this is due to the dynamical effect by the Alfv\'en waves. 
Note that 
the above value does not depend on the detailed state of the plasma, 
especially independent of the temperature on the star surface and in the magnetosphere. 
This is due to using the strong shock solution and the Rankine-Hugoniot relation, 
so that the above temperature can be obtained as long as we prepare conditions 
which are natural in the case of magnetar giant flares and can results in the initial strong shocks. 
As we will comment later, 
if we increase the temperature in the magnetosphere, 
the above strong shock condition is broken and the resulted maximum temperature decreases. 
%If we consider the propagation of the waves along a flux tube with length $R_{loop}$, 
%the time when the temperature start to rise 
%is the arriving time of the Alfv\'en waves at the loop top, 
%$t_{rise} \sim R_{loop} / c_A \sim R_{loop} / c$; 
%the peak time is the arriving time of the slow shock at the loop top, 
%$t_{peak} \sim R_{loop} / c_s = 3.3 R_{loop} / c$. 
%%If we assume the starting time of the initial spike is $t = 0$, 
%%which is equivalent to considering very short flux tubes, 
%The duration time can be written as $t_{peak} - t_{rise} \sim 2.3 R_{loop} / c$. 
%The observed duration time is $\sim$ 1 ms, 
%so that the necessary loop scale is about $R_{loop} \sim 10 R_0$. 
%Fig. \ref{fig:4.1.2} is the superposition of the temporal evolution of the temperature at the loop top 
%whose length is $R_{loop} = 2.5, 5, 7.5 R_0$. 
%Note that 
%each peak times deviates with respect to their length scale, 
%and the observed duration time may be the superposition of them. 

\citet{1995MNRAS.275..255T} showed that 
in the fireball region caused by a SGR flare phenomena 
the total Thomson optical depth across a scale of the neutron star radius $R_0 \sim 10^6$ [cm] becomes optically thick: $\tau_T \sim 5 \times 10^7 \gg 1$. 
The above estimate uses the typical values of the SGR flares, 
so the optical depth becomes also much larger than unity in the case of the magnetar giant flare.  
This means that 
the photons trapped in the fireball region cannot escape and cannot be observed. 
Our flare mechanism always works at the top of the magnetic loop, 
so that the generated photons can escape from the fireball region. 
In addition, 
because of the very short mean free path of the scattering between photons and pairs, 
the photon temperature takes the same value of the pair plasma. 
Hence, 
the observed photon temperature behaves as Figure \ref{fig:4.1.2}. 
As shown in Section \ref{sec:sec3.2}, 
the strength of the initially triggered slow shock gradually decreases 
after it reaches $r_{\mathrm{min}}$ given in Equation (\ref{eq:4.2}). 
Then, 
the spike form of the light curve gradually disappear during $\sim$ ms time scale, 
and at a later time we will observe the temperature on the surface of the fireball, 
as is indicated by observations~\citep{2006csxs.book..547W,2008A&ARv..15..225M}. 

In this paper, 
we assumed the mirror symmetry of flux tubes. 
This may be a too idealized situation for real magnetar giant flare, 
and imbalanced collision of the two slow shocks may be more realistic. 
However, as is discussed in Section \ref{sec:sec3.2}, 
the shocks become sufficiently strong if the flux tube length is longer than the critical length. 
Hence, we expect that even imbalanced collisions can produce sufficient energy as long as the collision occurred further than the critical length.

%\textcolor{red}{
Finally, 
we give a comment on a possibility of our shock collision mechanism playing an important role even in the twisted magnetic flux rope model. 
The collisionless magnetic reconnection converts the upstream magnetic field energy mainly into the plasma kinetic energy due to the collisionless nature, 
not in the thermal energy, 
so that another energy conversion mechanism is necessary for explaining the light curve of the giant flares in magnetars. 
We expect that our mechanism can give an explanation of the energy conversion from the kinetic energy into the thermal energy. 
%}

%\subsection{\label{sec:sec4.2}Open Flux Tube and Plasma Density}
%\subsection{\label{sec:sec4.2}Properties of Open Flux Tube}
\subsection{\label{sec:sec4.2}Properties of Closed Flux Tubes}

In this paper, 
we used 1-dimensional approximation 
%We consider this becomes a good approximation 
%because the strong magnetic field in the magnetar magnetosphere restricts the plasma motion along the magnetic field flux tube. 
and introduced the closed flux tube that super-radially expands, given by Equation (\ref{eq:2}). 
In this section, 
we discuss properties of the flux tube in magnetars. 
%More detailed dependence of numerical results on the configuration of the flux tube is discussed in Appendix \ref{sec:secA3}. 
Detailed dependence of numerical results on the configuration of the radially-expanding closed loop is given in Appendix C. %\ref{sec:secA3}. 

%%% funnel structure
This flux tube expands super-radially at the bottom. 
In the case of the Sun, 
this structure is considered to result from the rapid decrease of the gas pressure on the surface 
~\citep{1986SoPh..103..299S,2010NewAR..54...13T}. 
Inside of the Sun, 
there is a lot of hot matter, 
%and the gas pressure and the magnetic pressure are in nearly equipartition, 
and the gas pressure dominates the magnetic pressure, 
so that the flux tube of the magnetic field cannot expand. 
However, as leaving from the surface, 
the amount of matter and the gas pressure rapidly decrease 
and this allows the super-radial expansion of the flux tube due to its strong magnetic pressure. 
%Though there is no observational evidence of the existence of such a structure on the magnetar surface unfortunately, 
%we expect that 
%magnetars have such a structure between the plasma in the magnetosphere and atmosphere. %the crust region melted due to the strong heating of 
%some trigger phenomena of giant flares. 

%%% monopole like B
As indicated in Equation (\ref{eq:1}), 
the flux tube results in the split monopole like magnetic field $B \propto 1/r^2$ instead of the dipole field $B \propto 1/r^3$. 
In case of the Sun, 
the monopole like magnetic field results from the dragging of magnetic field line by the matter on the line. 
%which is accelerated outwards due the centrifugal force. 
On the other hand, 
the above mechanism does not seem to work on the magnetic field close to the magnetar surface 
because of the very strong magnetic field. 
However, 
in case of the giant flare, 
the observed isotropic peak luminosity of initial spike reaches $F_{\mathrm{SGR}} \sim 10^{46}$ [erg/s]
\citep{2005Natur.434.1110T,2005Natur.434.1098H,2007ApJ...665L..55T}, 
so that the energy density of the fireball region can be estimated as: 
\begin{eqnarray}
  \label{eq:4.2.1}
  e_{\mathrm{Spike}} &\sim& F_{\mathrm{Spike}} \Delta t / 4 \pi R^2 \Delta R 
  \nonumber
  \\
  &\sim& 10^{24} \left( \frac{\Delta t}{1 [\textrm{ms}]} \right)
  \left( \frac{R}{10^6 [\textrm{cm}]} \right)^{-2} 
  \nonumber
  \\
  &\times& \left( \frac{\Delta R}{10^6 [\textrm{cm}]} \right)^{-1} [\textrm{erg/cm$^3$}]
  ,
\end{eqnarray}
where $\Delta t$ is the duration of the initial spike, 
%$\Delta R$ is the radial coordinate of the edge of the fireball region. 
$\Delta R$ is the radius of the fireball region. 
The magnetic field energy density can be written as: 
\begin{equation}
  \label{eq:4.2.2}
  e_B = B^2/8 \pi \sim 10^{26} \left( \frac{B_0}{10^{14} [\textrm{G}]} \right)^2
  \left( \frac{R}{10^6 [\textrm{cm}]} \right)^{-6} [\textrm{erg/cm$^3$}]
  ,
\end{equation}
where $B_0$ is the magnetic field strength on the magnetar surface. 
Comparing Equations (\ref{eq:4.2.1}) and (\ref{eq:4.2.2}), 
the fireball energy density can exceed the magnetic field energy 
%as the radial coordinate increases sufficiently. 
when the fireball region expands sufficiently. 
Thus, 
we expect that 
the magnetic field can be dragged by the matter energy in the fireball region 
as the fireball evolves, 
and finally induces spherically expanding flux tube structure
~\footnote{
\citet{2005Natur.437..845F} discussed that 
the fireball region should be confined within a distance $\Delta R \sim 10 [\textrm{km}] \sim R_0$ 
to explain the luminosity and lifetime of the tail of the light curve; 
However, 
it might depend on the state of the plasma around the surface, 
so that in this paper we assume a fireball expanding more than the above $\Delta R$ like other existing works, 
such as \citep{2013ApJ...774...92P}. 
}
. 
Observationally, 
detection of large flux changes in the persistent X-ray flux of magnetars have been reported during giant flares~\citep{2001ApJ...552..748W}, 
and this is thought to be an indication of the change of the global magnetic field structure in magnetar magnetospheres. 
We consider that 
this is an evidence for occurring a similar phenomena reported in solar atmosphere, 
e.g. \citep{1986SoPh..103..299S,2010NewAR..54...13T}, 
which is also an change of magnetic field structure during solar flares, inducing super-radially expanding flux tubes. 
Although this indicates the highly dynamical flux tubes, 
we expect that the quasi-steady assumption of the flux tube is still not so bad approximation 
since the timescale of the evolution of flux tubes is thought to be comparable to that of tearing instability, 
which is longer than the characteristic timescale of our model, that is, the Alfv\'en crossing time. 
Note that 
even if the actual magnetic field is not the super-radial flux tube but the dipole, 
our flare mechanism can still be applicable 
since our mechanism does not depend strongly on the magnetic field structure 
as explained in Section \ref{sec:sec3.2}. 
In particular, 
even if we consider a flux tube with $f(r) \propto r^n$, 
the characteristic length-scale $r_{\mathrm{min}}$ given by Equation (\ref{eq:4.2}) 
%which determines the characteristic timescale 
becomes just $R_0^2 / n H$, 
and the characteristic timescale Equation (\ref{eq:4.3}) is nearly independent of the background magnetic field 
with low multi-pole components. 
Note that 
this argument can generally be used if the background plasma density is determined by the gravity of the central star. 
%In particular, 
%the characteristic lengthscale $r_{min}$ given by Equation \ref{eq:4.2} 
%which determines the characteristic timescale 
%is independent of the background magnetic field 
%and can be used if the background plasma density is determined by the gravity of the central star. 

Concerning another models of plasma density in magnetar magnetospheres, 
\citet{1995MNRAS.275..255T} estimated the plasma density in the fireball region in the case of $B \gg B_{\mathrm{QED}}$ and $k_B T \ll m_e c^2$ 
which is given as follows
\begin{equation}
  \label{eq:4.2.3}
  n^{\pm} = \frac{(m_e c)^3}{\hbar (2 \pi^3)^{1/2}} \left( \frac{B}{B_{\mathrm{QED}}} \right) \left( \frac{k_B T}{m_e c^2} \right) \exp \left[ - \frac{m_e c^2}{k_B T} \right]
  , 
\end{equation}
where $B_{\mathrm{QED}} = m_e^2 c^3 / e \hbar = 4.4 \times 10^{13}$[G] is the magnetic field strength at which the quantum effect starts to work. 
In the paper, 
they did not take into account the effect of the gravity. 
In addition, 
\citet{2002ApJ...574..332T} derived another form of the mass density 
in the case of the magnetosphere with toroidal magnetic field. 
To determine which model of the plasma density is the correct one is beyond the scope of our paper, 
and we would like to consider it for our future work. 

\subsection{\label{sec:sec4.3}Effects of Alfv\'en waves and $\sigma$-Parameter}

In this paper, 
we assume the continuous injection of Alfv\'en waves driven on the star surface. 
We found that 
the injected Alfv\'en waves transfer matter and energy through their propagation and dissipation, 
and they also play a key role to the flare phenomena 
through heating the initially driven slow shock by their dissipation. 
The Alfv\'en wave energy density can be written as follows:
\begin{equation}
  \label{eq:5.1}
  e_{\mathrm{A,wave}} = (\delta B)^2/8 \pi = (1/\sigma + 1) (\delta v_{\perp}/c)^2 e_B
  ,
\end{equation}
where $e_B = B^2 / 8 \pi$ is the background magnetic field energy density 
and we used Equation (\ref{eq:B1}) and $\sigma = B^2/ 4 \pi \rho h$. 
The energy flux of Alfv\'en waves can be written as: 
\begin{eqnarray}
  \label{eq:5.2}
  F_{\mathrm{A,wave}} &=& e_{\mathrm{A,wave}} R^2 v_A 
  \nonumber
  \\
%  &\sim& 10^{49} \left( \frac{\delta v_{\perp}/c}{\textcolor{blue}{0.75 /\sqrt{3}}} \right)^2 \left( \frac{c_A}{3 \times 10^{10} [\textrm{cm /s}]} \right) 
%  &\sim& 10^{49} \left( \frac{\delta v_{\perp}/c}{0.75 c_{s,r}/c} \right)^2 \left( \frac{c_A}{3 \times 10^{10} [\textrm{cm /s}]} \right) 
  &\sim& 10^{45} \left( \frac{\delta v_{\perp} / c}{ 10^8 \mathrm{[cm/s]}/c } \right)^2 \left( \frac{c_A}{3 \times 10^{10} [\textrm{cm/s}]} \right) 
  \nonumber
  \\
  && \left( \frac{B_0}{10^{14} [\textrm{G}]} \right)^2 \left( \frac{R}{10^{6} [\textrm{cm}]} \right)^2 
  [\textrm{erg/s}]
  .
\end{eqnarray}
%In this calculation, 
%we used typically $\delta v_{\perp} = 0.75 c / \sqrt{3}$. 
%where $c_{s,r} \equiv c / \sqrt{3}$ is the relativistic sound speed. 
where $\delta v_{\perp} = 10^8$[cm/s] is a typical shear wave velocity in crusts obtained by \citet{1989ApJ...343..839B}. 
Equation (\ref{eq:5.2}) indicates that 
%the above amplitude is too large to explain the observation. 
the amplitude used in our simulations, $\langle \delta v_{\perp} \rangle = 0.75 c_{s,r}$, is too large to explain the observation. 
However, 
as discussed in Section \ref{sec:sec3.2}, 
the most important effect of the Alfv\'en waves is to supply energy to the initially triggered slow shock 
and keep its behavior to satisfy the strong shock condition. 
If we assume that a fraction $\epsilon$ of the Alfv\'en wave energy density dissipate into thermal energy, 
the resulted temperature can be expressed as $k_B T / m c^2 = \epsilon e_{\mathrm{A,wave}} / \rho c^2 \sim \epsilon \sigma (\delta v_{\perp}/c)^2$. 
Using typical parameters for magnetars, $\langle \delta v_{\perp} \rangle / c \sim 10^{-2}$~\citep{1989ApJ...343..839B} and $\sigma \sim 10^4$ (see also the footnote 
\footnote{
In particular, 
even if half of the initial spike energy of a giant flare could be converted into the particle energy, 
the $\sigma$ value in the magnetar magnetosphere is  
$\sigma_{\mathrm{SGR}} = (B^2 c / 4 \pi) / (F_{\mathrm{SGR}} / 4 \pi R_0^2) \sim 10^4$ 
where $F_{\mathrm{SGR}} \sim 10^{46}$ [erg/s] is the isotropic peak luminosity of initial spike of a giant flare~
\citep{2005Natur.434.1110T,2005Natur.434.1098H,2007ApJ...665L..55T}, 
$B \sim 10^{14}$ [G] is the magnetic field strength at the surface 
and $R_0 \sim 10^6$ [cm] is the neutron star radius. 
Hence, the $\sigma$ value in magnetar magnetospheres might be at least $10^4$. 
}
)
,
even small dissipation, such as $\epsilon \sim 10^{-2}$, is sufficient for resulting in heating the magnetosphere more than $50$ [keV], 
which is enough to keep the strength of the slow shock. 
In Appendix B, %\ref{sec:secA.2}, 
we also provide a study of behaviors of Alfv\'en waves in Poynting-dominated plasma with $\sigma$ value from $1$ to $500$. 
The results show that 
though the dissipation rate becomes smaller as the $\sigma$ parameter increases, 
the resulting temperatures are almost independent of the $\sigma$ parameter 
due to the increase of the magnetic field energy comparing with the thermal energy of the background plasma. 
This indicates that 
the increase of $\sigma$ parameter in a magnetar magnetosphere makes our flare mechanism easier to work. 

In this paper, 
we also assumed the MHD approximation in the magnetar magnetosphere. 
%Differently from the pulsar magnetosphere, such as the Crab pulsar, 
%there might be no gap and no pair creation might occur in the magnetar magnetosphere, 
%so the MHD approximation is also doubtful in the quasi steady state. 
%However, 
If some burst phenomena occur and result in a fireball, 
the number density of the electron positron pair increases drastically. 
In addition, 
the rapid interaction between photons and pairs allows the particle distribution to be an equilibrium state. 
Hence, 
we expect that 
at least in the fireball region we can use MHD approximation 
and Alfv\'en waves can propagate. 
Concerning the dissipation of the Alfv\'en waves, 
we considered the dissipation through the shock forming by the non-linear effect of waves. 
However, 
there are other possibilities. 
For example, 
the wave decay process can play an important role 
since the particle number density is small in the vicinity of the shock front of the initially triggered slow shock. 
Recently, 
it was proved that the wave decay processes can dissipate Poynting energy in a back ground plasma 
very efficiently~\citep{2013ApJ...770...18A,2013ApJ...771...53M,2013arXiv1303.6781M}. 
Since the main role of Alfv\'en waves in our mechanism is the heating of the slow shock, 
this means that 
the plasma effects on the decay of the Alfv\'en waves do not change our physical picture 
but even encourages to work.

\acknowledgments

We would like to thank John Kirk, Iwona Mochol, Simone Giacche, Kohta Murase, and Susumu Inoue,  
for many fruitful comments and discussions. 
We also would like to thank our anonymous referee for a lot of fruitful comments on our paper. 
%Numerical computations were carried out 
%on SR16000 at YITP in Kyoto University. 
%Calculations were also carried out on the Cray XT4 
%at Center for Computational Astrophysics, CfCA, of National Astronomical Observatory of Japan.
This work is supported by Max-Planck-Institut f\"ur Kernphysik. 
This work is supported in part by the Postdoctoral Fellowships for Research Abroad program by the Japan Society for the Promotion of Science No. 20130253 (M. T.). 
This work is also supported in part by Grants-in-Aid for 
Scientific Research from the MEXT of Japan, 22864006 (TKS), and 24540258 (TT), 
and also by Grant-in-Aid for Scientific Research on Innovative Areas No. 24103006 (SK). 
%SR16000'à


%% in parentheses, after the keyword, but they will not be verified.

%%{\it Facilities:} \facility{Nickel}, \facility{HST (STIS)}, \facility{CXO (ASIS)}.



\appendix

\section{\label{sec:secA.1}Numerical Method}

In this appendix, 
we present the detailed explanation of the numerical method 
we use in this paper. 
Variables indicated by Greek letters take values from $0$ to $3$, 
and those indicated by Roman letters take values from $1$ to $3$.

As is mentioned in Sec. \ref{sec:sec2.2}, 
we solve the dynamical evolution of the plasma and waves by solving the ideal GRMHD equations 
in Schwarzschild spacetime 
whose 
\begin{eqnarray}
  \label{eq:A.1}
  ds^2 &=& g_{\mu \nu} d x^{\mu} dx^{\nu} 
  , 
  \\
  &=& - w(r) dt^2 + w(r)^{-1} dr^2 + r^2 (d \theta^2 + \sin^2 \theta d \phi^2)
  ,
  \nonumber
\end{eqnarray}
where $w(r) = 1 - r_g / r$, 
$r_g \equiv 2 M G / c^2$ is the Schwarzschild radius, 
$M, G, c$ are the central star mass, the gravitational constant and the light velocity, respectively. 
We use the metric signature, $(-,+,+,+)$, along with units $c = 1$. 

The basic equations are the mass conservation, the energy-momentum conservation and the induction equation 
which are given as:
\begin{eqnarray}
  \label{eq:A.1}
   \nabla_{\mu} N^{\mu} &=& \frac{1}{\sqrt{- g}} \partial_{\mu} [\sqrt{- g} N^{\mu}] = 0
   ,
   \\
  \label{eq:A.2}
   \nabla_{\mu} T^{\mu}_{\nu} &=& \frac{1}{\sqrt{- g}} \partial_{\mu} [\sqrt{-g} T^{\mu}_{\nu}] - \frac{1}{2} g_{\mu \rho, \nu} T^{\mu \rho} = 0
   ,
   \\
  \label{eq:A.3}
  \nabla_{\mu} (^*F^{\mu \nu}) &=& \frac{1}{\sqrt{- g}} \partial_{\mu} [\sqrt{-g} (^*F^{\mu \nu})] = 0
  ,
\end{eqnarray}
where $N^{\mu} \equiv \rho u^{\mu}$ is the mass flux vector, 
$\rho$ is the mass density in the fluid comoving frame, 
$u^{\mu}$ is the 4-velocity 
which satisfies $u^{\mu} u_{\mu} = - 1$; 
$T^{\mu \nu}$ is the energy-momentum tensor defined as 
\begin{equation}
  \label{eq:A.4}
   T^{\mu \nu} = (\rho h  + |b|^2) u^{\mu} u^{\nu} + (p_{\mathrm{gas}} + |b^2| / 2) g^{\mu \nu} - b^{\mu} b^{\nu}  
   ,
\end{equation}
where $h$ is the specific enthalpy, 
$p$ is the gas pressure, 
$b^{\mu} = (g_{i \mu} B^i u^{\mu}, (B^i + u^i b^t) / u^t)$ is the magnetic field 4-vector, 
$B^i$ is the magnetic field in the laboratory frame; 
$^*F^{\mu \nu}$ is the dual of the electromagnetic field strength tensor 
given by $^*F^{\mu \nu} = u^{\mu} b^{\nu} - u^{\nu} b^{\mu}$. 

To use the HLLD approximation, 
we introduce the following 4-vector $\tilde{u}^{\mu} = (\tilde{\gamma}, \tilde{\gamma} \tilde{ {\bf v} })$ 
\begin{equation}
  \label{eq:A.5}
  \tilde{u}^{\mu} \equiv (\sqrt{w} u^t, u^r / \sqrt{w}, r u^{\theta}, r \sin \theta u^{\phi})
  .
\end{equation}
Using this 4-vector, 
the square of the 4-velocity $u^{\mu}$ can be expressed as: $u_{\mu} u^{\mu} = - (\tilde{u}^t)^2 + (\tilde{u}^r)^2 + (\tilde{u}^{\theta})^2 + (\tilde{u}^{\phi})^2 
= \eta_{\mu \nu} \tilde{u}^{\mu} \tilde{u}^{\nu} = -1$ 
where $\eta_{\mu \nu} = \mathrm{diag} (-1,1,1,1)$. 
In addition, 
we also introduce the following 3-vector $\tilde{B}^i$ 
\begin{equation}
  \label{eq:A.5}
  \tilde{B}^i = (B^r, \sqrt{w} r B^{\theta}, \sqrt{w} r \sin \theta B^{\phi})
  .
\end{equation}
Using this 3-vector, 
the 4-magnetic field vector $b^{\mu}$ can be expressed as follows: 
\begin{equation}
  \label{eq:A.6}
  b^{\mu} = (\tilde{b}^t / \sqrt{w}, \sqrt{w} \tilde{b}^r, \tilde{b}^{\theta} / r, \tilde{b}^{\phi} / r \sin \theta)
\end{equation}
where $\tilde{b}^{\mu}$ is defined as: 
\begin{equation}
  \label{eq:A.7}
  \tilde{b}^{\mu} = (\tilde{\gamma} (\tilde{{\bf B}} \cdot \tilde{ {\bf v} }), 
  \quad \tilde{{\bf B}} / \tilde{\gamma} + \tilde{\gamma} (\tilde{{\bf B}} \cdot \tilde{ {\bf v} }) \tilde{ {\bf v} } )
  ,
\end{equation}
where $\tilde{ {\bf B} } \cdot \tilde{ {\bf v} } = \tilde{B}^r \tilde{v}^r + \tilde{B}^{\theta}\tilde{v}^{\theta} + \tilde{B}^{\phi} \tilde{v}^{\phi}$. 
Similarly to $\tilde{u}^{\mu}$, 
the square of the 4-magnetic field vector can be rewritten as: 
$b_{\mu} b^{\mu} = \eta_{\mu \nu} \tilde{b}^{\mu} \tilde{b}^{\nu}$. 

Using these new vectors, 
the basic equations Equations (\ref{eq:A.1}), (\ref{eq:A.2}), (\ref{eq:A.3}) can be rewritten as 
\begin{eqnarray}
  \label{eq:A.7}
  \partial_t U + \partial_r F^r + S = 0
  ,
  \\
  \tilde{B}^r \sqrt{-g} = \mathrm{const.}
  ,
\end{eqnarray}
where
\begin{eqnarray}
U = 
\left(
 \begin{array}{ccc}
   \sqrt{- g / w} &\times& \rho \tilde{\gamma}\\
   \sqrt{-g} / w &\times& [(\rho h + \tilde{b}^2) \tilde{\gamma}^2 \tilde{v}^r - \tilde{b}^t \tilde{b}^r ] \\
   r \sqrt{-g} / \sqrt{w} &\times& [(\rho h + \tilde{b}^2) \tilde{\gamma}^2 \tilde{v}^{\perp} - \tilde{b}^t \tilde{b}^{\perp} ] \\
   \sqrt{-g} &\times& [(\rho h + \tilde{b}^2) \tilde{\gamma}^2 - (p_{\mathrm{gas}} + \tilde{b}^2 / 2) - (\tilde{b}^t)^2 ] \\
   \sqrt{- g / w} / r &\times& \tilde{B}^{\perp}
 \end{array}
\right), 
\end{eqnarray}
\begin{eqnarray}
F^r = 
\left(
 \begin{array}{ccc}
   \sqrt{- g  w} &\times& \rho \tilde{\gamma} \tilde{v}^r \\
   \sqrt{-g} &\times& [(\rho h + \tilde{b}^2) \tilde{\gamma}^2 (\tilde{v}^r)^2 + (p_{\mathrm{gas}} + \tilde{b}^2 / 2) - (\tilde{b}^r)^2 ] \\
   r \sqrt{-g w} &\times& [(\rho h + \tilde{b}^2) \tilde{\gamma}^2 \tilde{v}^{r} \tilde{v}^{\perp} - \tilde{b}^r \tilde{b}^{\perp} ] \\
   w \sqrt{-g} &\times& [(\rho h + \tilde{b}^2) \tilde{\gamma}^2 \tilde{v}^r - \tilde{b}^t \tilde{b}^r ] \\
   \sqrt{- g w} / r &\times& [\tilde{B}^{\perp} \tilde{v}^r - \tilde{B}^r \tilde{v}^{\perp} ]
 \end{array}
\right), 
\end{eqnarray}
\begin{eqnarray}
S = 
\left(
 \begin{array}{c}
   0\\
   -\frac{\sqrt{- g}}{2} \left[ \frac{(r^2)'}{r^2} \left\{ (\rho h + \tilde{b}^2) (\tilde{u}^{\perp})^2 + 2 (p_{\mathrm{gas}} + \tilde{b}^2 / 2) - (\tilde{b}^{\perp})^2 \right\} \right.
   \\ 
   - \left. \frac{w'}{w} \left\{ (\rho h + \tilde{b}^2) [\tilde{\gamma}^2 + (\tilde{u}^r)^2] - (\tilde{b}^t)^2 - (\tilde{b}^r)^2 \right\}\right] \\
   0 \\
   0 \\
   0
 \end{array}
\right), 
\end{eqnarray}
where $\perp = \theta, \phi$. 
Note that the conservative variables $U$ and the numerical flux $F^r$ have a form $\alpha_{\mathrm{curv}} U_{\mathrm{RMHD}}$ and $\beta_{\mathrm{curv}} F_{\mathrm{RMHD}}$ 
where $U_{\mathrm{RMHD}}$ and $F_{\mathrm{RMHD}}$ are the conservative variables and the numerical flux of the relativistic MHD equations in the flat geometry, 
$\alpha_{\mathrm{curv}}$ and $\beta_{\mathrm{curv}}$ are curvature terms. 
We calculate numerical flux $F^r$ as the following form, $F^r = \beta_{\mathrm{curv}} F_{\mathrm{HLLD}}$, 
where $F_{\mathrm{HLLD}}$ is the HLLD flux calculated using $U_{\mathrm{RMHD}}$ and $F_{\mathrm{RMHD}}$. 
In this case, 
the HLLD flux is not a good approximation comparing with the flat geometry case. 
However, 
it is a better estimation for the numerical flux, 
especially when $\Delta r / r \ll 1$ where $\Delta r$ is the mesh size. 
Concerning the term $S$, 
we calculate it as a source term at $t = t^n + \Delta t / 2$ 
where $t_n$ is the time at n-th time steps 
and $\Delta t$ is the size of a time step. 
To calculate the primitive variables from the conservative variables, 
we used a method developed by \citet{2007MNRAS.378.1118M}
which allows us to calculate the primitive valuables very accurately. 
(see also \citep{2011ApJS..193....6B})
Note that we calculate along flux tube with super-radial expansion structure, 
so that the radial coordinate $r$ accompanying with the angular components $(\theta, \phi)$ of the above tensors 
should be changed into $r \sqrt{f}$ 
where $f$ is defined in Equation (\ref{eq:2}), 
for example, $\sqrt{-g} \equiv r^2 f \sin \theta$.

\section{\label{sec:secA.2}Propagation of Alfv\'en Waves in Poynting-Dominated Plasmas}

In this section, 
we investigate the $\sigma$-dependence of an evolution of the linearly polarized Alfv\'en waves with large amplitude. 
Differently from the circularly polarized case, 
the linearly polarized Alfv\'en wave with large amplitude is not an exact solution of RMHD equations. 
Hence, 
immediately after their generation, e.g. by an oscillation of a neutron star surface, 
they change the background plasma structure by their magnetic pressure force 
and it induces their effective dissipation. 

\begin{figure}[h]
 \centering
  \includegraphics[width=6.5cm,clip]{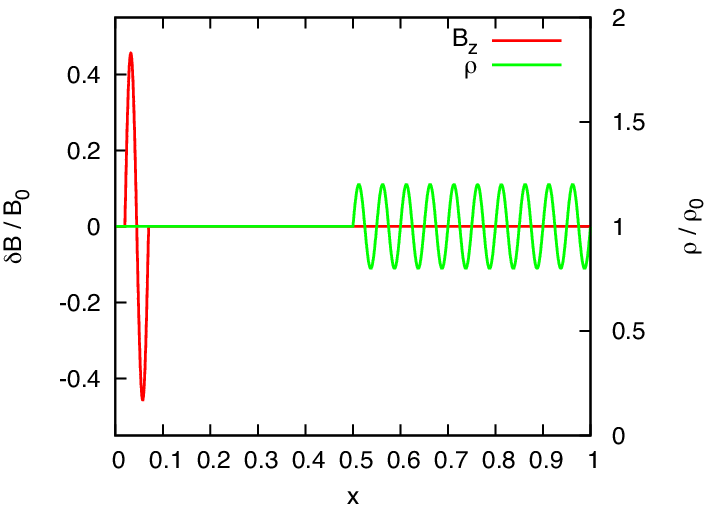}
  \includegraphics[width=6.5cm,clip]{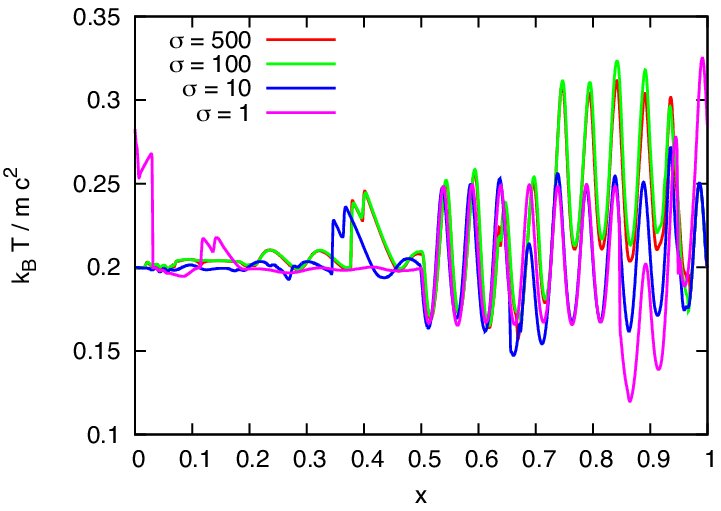}
  \includegraphics[width=6.5cm,clip]{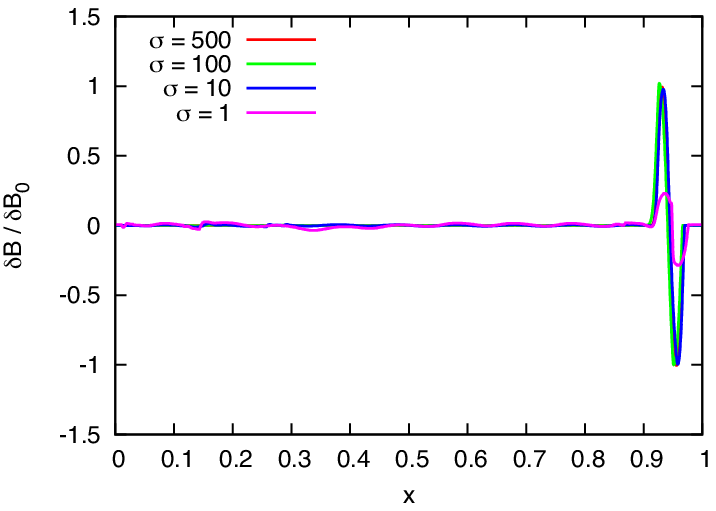}
  \includegraphics[width=6.5cm,clip]{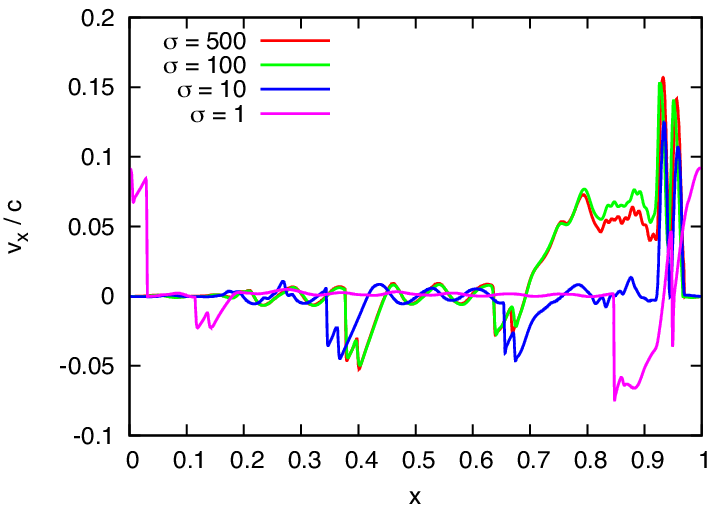}
  \caption{Snapshots of the propagation of an Alfv\'en wave. 
           Left top: initial snapshots of the density and magnetic field. 
           Left bottom: snapshot of the magnetic field in the unit of the initial amplitude at $0.9$ Alfv\'en crossing time. 
           Right top: snapshot of the temperature at $0.9$ Alfv\'en crossing time. 
           Right bottom: snapshot of the velocity $v_x$ at $0.9$ Alfv\'en crossing time. 
          }
  \label{fig:A2.1}
\end{figure}

In this calculation, 
we prepared a numerical domain: $0 < x < 1$, 
and divided it using uniform meshes with size $\Delta x = 2.5 \times 10^{-4}$. 
The background temperature is $p_0 / \rho_0 = 0.2$ 
where subscript $0$ means the variables are averaged values. 
We used the ideal equation of state $h = 1 + \Gamma / (\Gamma - 1) p / \rho$ with $\Gamma = 5 / 3$. 
For the Alfv\'en wave, 
we prepared a linearly polarized Alfv\'en wave with amplitude $\delta B = B_0 / 2$ 
whose wavelength is $0.05$ times of the length of the numerical domain. 
To simulate disturbances from the density decrease in the magnetar magnetosphere, 
we set sinusoidal density disturbance with amplitude $0.2 \rho_0$ as in the left-top panel of Figure \ref{fig:A2.1}. 
We used the periodic boundary condition. 
For the initial condition of an Alfv\'en wave, 
we consider 
\begin{equation}
B^z = \delta B \sin(2 \pi x / \lambda)
, \quad
v^z = - \frac{B^z}{\sqrt{4 \pi \rho h + |B|^2}}
, 
\label{eq:B1}
\end{equation}
where $\lambda = 1/20$. 
Other vector variables are set to $0$. 

Left-bottom panel of Figure \ref{fig:A2.1} is a snapshot of magnetic field $B^z$ at 0.9 Alfv\'en crossing time. 
We found that 
the Alfv\'en wave keeps its original form in high-$\sigma$ plasma case: $\sigma > 10$; 
on the other hand, 
in the case of $\sigma = 1$, 
the amplitude of the Alfv\'en wave became nearly a quarter of the initial value. 
This means 
the efficiency of dissipation of Alfv\'en wave energy increases with decreasing the magnetization parameter $\sigma$, 
and low-$\sigma$ Alfv\'en waves lose almost all of its energy after propagating approximately 20-wavelength. 
Right-top panel of Figure \ref{fig:A2.1} is a snapshot of the temperature $k_B T / m c^2$ at 0.9 Alfv\'en crossing time. 
This panel shows that 
the background plasma temperature becomes hotter due to the dissipation of Alfv\'en waves 
as increasing the magnetization parameter $\sigma$; 
and when $\sigma > 100$, 
the resulted temperature seems to reach a saturation value $\sim 0.3 m c^2$. 
%Concerning the induced temperature, 
%it is nearly proportional to $\sigma$ when $\sigma > 10$. 
This might seem to contradict the previous conclusion 
that the amount of dissipation of Alfv\'en waves decreases with increasing $\sigma$. 
However, this can be explained as follows. 
As $\sigma$ increases, 
the energy of Alfv\'en waves increases. 
This means that 
in high-$\sigma$ plasmas 
%the total dissipated energy of Alfv\'en waves can be very large 
the dissipated energy density of Alfv\'en waves per unit mass can be very large, 
even though their dissipation efficiency is very small 
because of their large energy.  
The right-bottom panel of Figure \ref{fig:A2.1} is a snapshot of the radial velocity at 0.9 Alfv\'en crossing time. 
Similarly to the temperature, 
the induced radial velocity by the Alfv\'en wave becomes larger as $\sigma$ increasing 
and seems to reach a saturation value $\sim 0.15 c$. 

From the above consideration, 
we can conclude that 
in the Poynting-dominated plasma 
the dissipation efficiency of linearly polarized Alfv\'en waves becomes small 
%but their total dissipation energy becomes large as $\sigma$ increasing; 
but their dissipation energy density becomes large as $\sigma$ increasing; 
the induced temperature and radial velocity reach a saturation values 
when $\sigma > 100$. 

Note that 
when we consider plasma with $\sigma > 1$ 
the change of the induced velocity and temperature are very small, 
typically just factor of a few. 
This indicates that 
the physics of Alfv\'en waves in a very high-$\sigma$ plasma 
can be approximated with considerable accuracy using a plasma with $\sigma > 10$; 
and our simulations with $\sigma = 10$ would be a good approximation of the evolution of Alfv\'en waves in a magnetar magnetosphere. 

\section{\label{sec:secA3}Effects of Configuration of Radially Expanding Closed Flux Tubes}

\begin{figure}[h]
 \centering
  \includegraphics[width=6.5cm,clip]{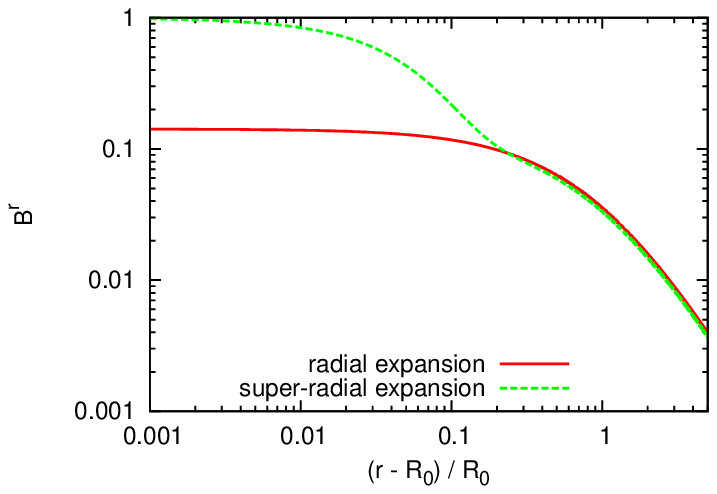}
  \includegraphics[width=6.5cm,clip]{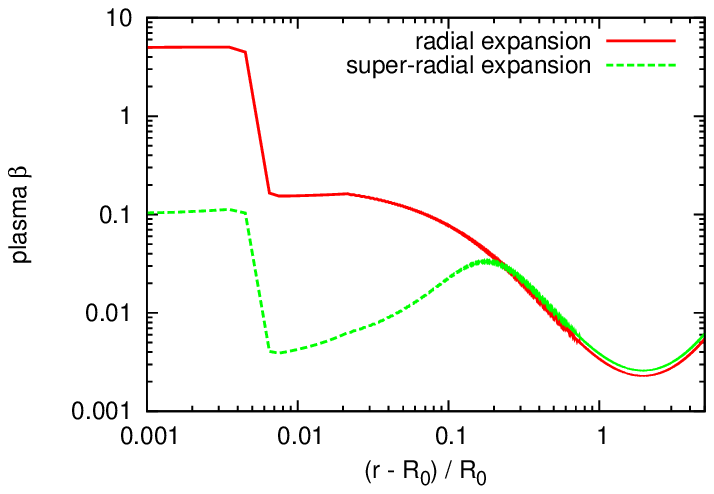}
  \caption{Snapshots of the profiles of $B^r$ and plasma $\beta$ %and the total pressure $p_{gas} + b^2 / 2$ 
           at the initial time. 
           The top panel is $B^r$ and the bottom panel is the plasma $\beta$. %and bottom panel is the total pressure. 
          }
  \label{fig:3.1.5.0}
\end{figure}

%In this paper, 
%we assume the existence of the funnel flux tube structure similar to the solar atmosphere case. 
In this paper, 
we adopt super-radially expanding flux tubes, referring to closed loops on the Sun. 
In the following, 
we check the effect of this flux tube on the evolution of the magnetosphere plasma. 
Note that 
Equation (\ref{eq:1}) means the funnel structure weakens the magnetic field around $r = R_1 = 2 R_0$. 
%Here we choose the strength of the magnetic field 
%as its value in the outer region agrees with the case including the flux tube structure, 
%which means the magnetic field at the surface of the case without the flux tube is weaker than 
%that of the case with the flux tube. 
%As a comparison 
We adopt a radially expanding flux tube with the same magnetic field strength at the loop top 
to compare with the case of the super-radially expanding tube 
we have considered so far: $f(r) = \mathrm{const.}, B^r \propto 1/r^2$. 
As shown in the top panel of Figure \ref{fig:3.1.5.0}, 
the magnetic field strength of the radially expanding flux tube case near the surface is smaller than that of the super-radial expansion case. 
This results in larger plasma $\beta$ near the surface. 
%and larger initial pressure gradient near the surface originated from the gas pressure gradient between the star atmosphere and the magnetosphere. 

\begin{figure}[t]
 \centering
  \includegraphics[width=5.3cm,clip]{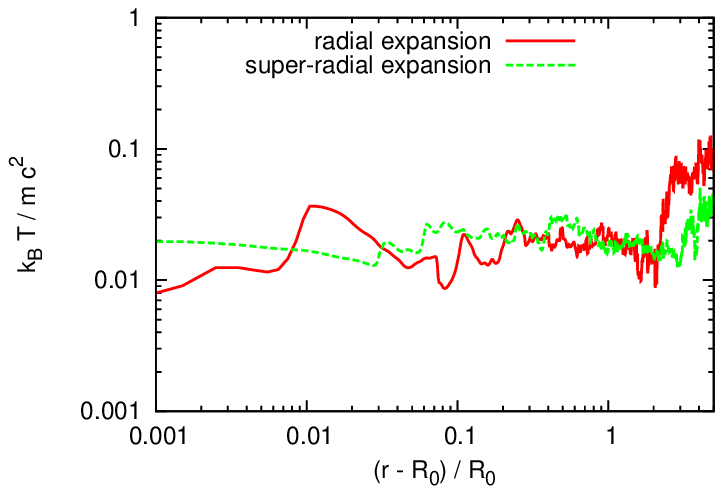}
  \includegraphics[width=5.3cm,clip]{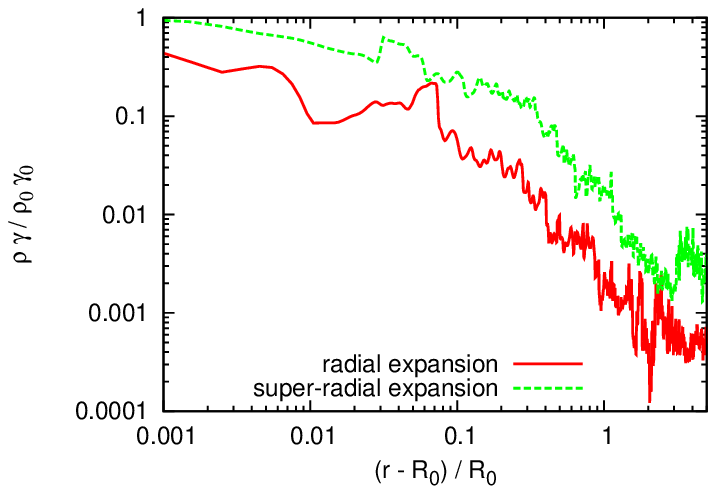}
  \includegraphics[width=5.3cm,clip]{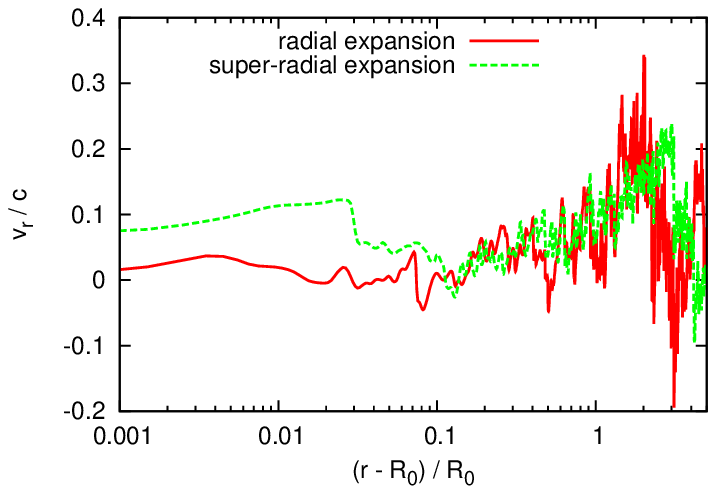}
  \caption{Top: A snapshot the temperature $k_B T / m c^2$. 
           Middle: A snapshot of the density $\rho \gamma / \rho_0 \gamma_0$. 
           Bottom: A snapshot of the radial velocity $v_r / c$.  
           All snapshots are at $t = 10 t_0$ 
           and we used $r_g / R_0 = 0.3$ and $\langle \delta v_{\perp} \rangle = 0.75 c_s$. 
          }
  \label{fig:3.1.5}
\end{figure}

Each panel of Figure \ref{fig:3.1.5} is the snapshots of the temperature, mass density and radial velocity %at the loop top 
in the case of $r_g / R_0 = 0.3$ and $\langle \delta v_{\perp} \rangle = 0.75 c_s$ at $t = 10 t_0$ 
comparing with the radially expanding flux tube case. 
%We find that 
%in the case without the flux tube structure 
%the mass and energy transfer are smaller comparing with the case including the flux tube structure, 
%though their behavior are very similar. 
We find that 
the radial expansion case gives smaller mass density. % and temperature. 
In particular, 
the radial velocity in the case of radially-expanding flux tube structure is clearly smaller 
in the region $r < R_1$. 
This is because 
the magnetic field with the radially-expanding flux tube is weaker in this region 
and this makes the Alfv\'en wave pressure smaller, 
which results in the smaller mass and energy transfer to the outer region. 
Note that 
the smaller mass density near the surface is induced by the aforementioned initial stronger pressure gradient force 
and does not relate to the Alfv\'en waves. 

Note that 
if we use the same magnetic field value at the central star surface, 
%the flux tube structure results in the weaker magnetic field in the outer region 
the super-radial expansion results in the weaker magnetic field in the outer region 
and this induces smaller Alfv\'en wave pressure and mass and energy transfer. 

\begin{figure}[t]
 \centering
  \includegraphics[width=5.3cm,clip]{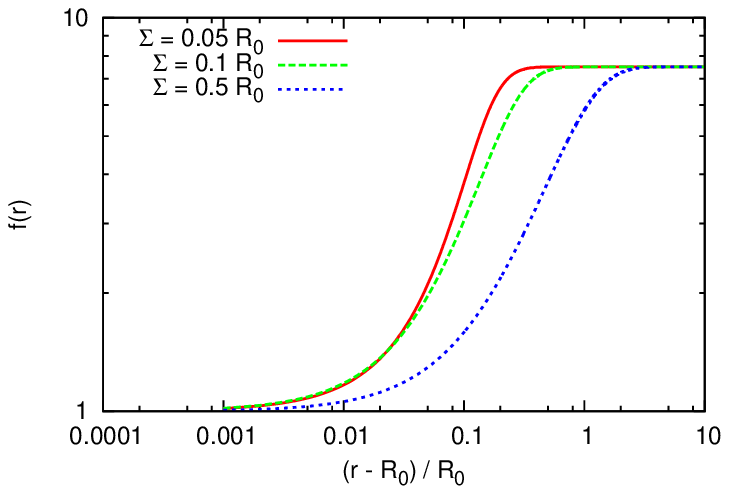}
  \includegraphics[width=5.3cm,clip]{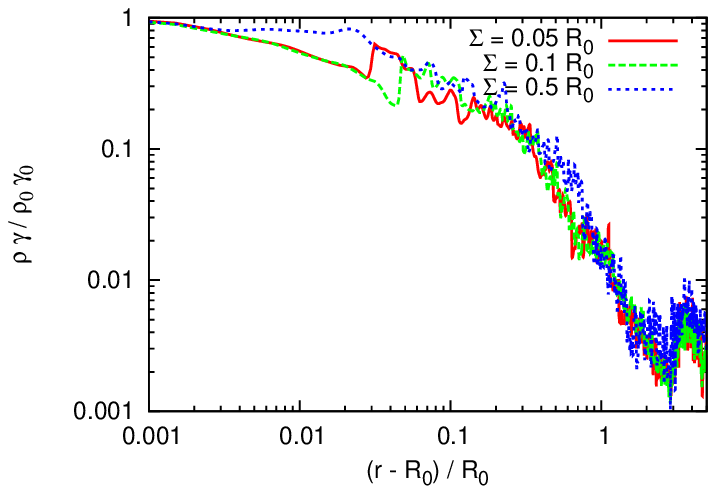}
  \includegraphics[width=5.3cm,clip]{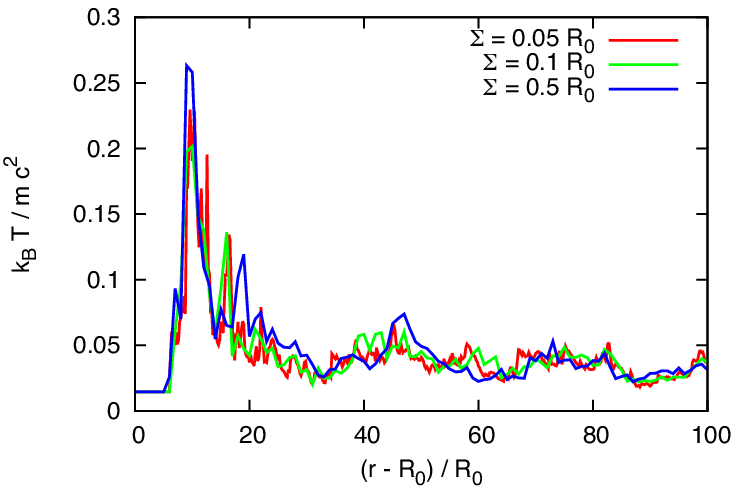}
  \caption{Left: The configuration of the flux tube with different value of $\Sigma$. 
           Middle: Snapshots of the density profile at $t = 10 R_0 / c$. 
           Right: The temporal evolution of temperature at the loop top. 
          }
  \label{fig:C.1}
\end{figure}

Finally, 
we discuss the dependence on the parameter $\Sigma$. 
Equation (\ref{eq:2}), has three parameters: $R_1, \Sigma, f_{\mathrm{Max}}$. 
$R_1$ controls the coordinate 
where the super-radial flux tube expands drastically. 
This should be determined by the pressure balance, 
so that we use the same coordinate as the edge of the atmosphere. 
$f_{\mathrm{Max}}$ determines the strength of the magnetic field in the magnetosphere, 
and we set this value through the numerical limitation of low-$\beta$ calculations. 
Concerning $\Sigma$, 
this parameter determines the characteristic scale of the expansion of the tube 
as can be seen in the left panel of Figure \ref{fig:C.1} 
that is the configuration of flux tubes with different $\Sigma$ values. 
The middle and right panels of Figure \ref{fig:C.1} are 
a snapshot of the density profile at $t = 10 R_0$ and the temporal evolution of temperature at the loop top 
using different values of $\Sigma$. 
Within the probable parameter region $\Sigma < R_1$, 
we cannot find any drastic differences in those results. 

%\subsection{Convergence Check}
%
%Fig. \ref{fig:3.1.4} is the temporal evolution of the time averaged mass flux at the loop top in the unit of the magnetic field energy density at the surface
%\begin{equation}
%  \langle \rho \gamma c^2 v_r / E_B c \rangle \equiv \frac{1}{T} \int^T_0 dt \frac{\rho \gamma c^2 v_r}{c B_0^2 / 8 \pi}
%  , 
%\end{equation}
%using different resolutions: $N = 2500, 5000, 7500$. 
%This figure shows that 
%the time averaged numerical results show a good convergence as increasing the mesh number. 
%Note that 
%the mass flux becomes larger as the resolution decreases in the initial phase. 
%This can be explained as follows. 
%The amplitude of the Alfv\'en waves decays faster 
%as the mesh number decreases due to the large numerical dissipation. 
%This means 
%the wave pressure, $P_w = \delta B_{\perp}^2 / 8 \pi$, decreases more rapidly with increasing the radial coordinate 
%when we use coarse meshes, 
%and this results in the larger wave pressure gradient force in the outwards direction 
%and larger mass flux. 
%
%\begin{figure}[h]
% \centering
%  \includegraphics[width=6.5cm,clip]{./graphics/convergence_rg30_amp075_vAM.eps}
%  \caption{The temporal evolution of the time averaged mass flux $\rho \gamma c^2 v_r / c E_B$ at the loop top. 
%           Calculations have done for different resolutions: $N = 2500, 5000, 7500$. 
%           The figure shows that the calculations shows a good convergence. 
%          }
%  \label{fig:3.1.4}
%\end{figure}

%\bibliography{ANS}

\end{document}